\documentclass[9pt,sigconf]{acmart}
\usepackage{amsmath}
\usepackage[linesnumbered,lined,boxed,commentsnumbered,ruled,vlined]{algorithm2e}
\usepackage{graphicx}
\usepackage{subcaption}
\usepackage{wrapfig}
\usepackage{multirow}
\usepackage{soul}
\usepackage{stmaryrd}

\AtBeginDocument{%
	\providecommand\BibTeX{{%
			\normalfont B\kern-0.5em{\scshape i\kern-0.25em b}\kern-0.8em\TeX}}}

\setcopyright{acmcopyright}
\copyrightyear{2018}
\acmYear{2018}
\acmDOI{10.1145/1122445.1122456}

\acmConference[Woodstock '18]{Woodstock '18: ACM Symposium on Neural
	Gaze Detection}{June 03--05, 2018}{Woodstock, NY}
\acmBooktitle{Woodstock '18: ACM Symposium on Neural Gaze Detection,
	June 03--05, 2018, Woodstock, NY}
\acmPrice{15.00}
\acmISBN{978-1-4503-XXXX-X/18/06}

\newcommand{\eat}[1]{}
\newcommand{\itvl}[1]{\llbracket #1 \rrbracket}
\newcommand{\itvllo}[1]{\llparenthesis #1 \rrbracket}

\begin{document}
	
	\title{{\indextitle}: A (G)eneric (L)earned (In)dexing Mechanism for Complex Geometries}
	
	
	
	\author{Congying Wang}
	\authornote{Part of this work was conducted while the author was affiliated with WSU.}
	\affiliation{%
		  \institution{University at Buffalo}
		  \country{}
		}
	\email{cwang39@buffalo.edu}
	
	\author{Jia Yu}
	\affiliation{%
		  \institution{WSU, Wherobots Inc.}
		  \country{}
		}
	\email{jia.yu1@wsu.edu}
	
	\author{Zhuoyue Zhao}
	\affiliation{%
		\institution{University at Buffalo}
		\country{}
	}
	\email{zzhao35@buffalo.edu}
	
	\newtheorem{defn}{definition}
	\AfterEndEnvironment{defn}{\ignorespaces}
	\newtheorem{theo}{theorem}
	\AfterEndEnvironment{theo}{\ignorespaces}
	\newtheorem{lem}{lemma}
	\AfterEndEnvironment{lem}{\ignorespaces}
	
	\newcommand{\indextitle}{\textsc{GLIN}}
	\newcommand{\rangename}{\text{interval}}
	\newcommand{\geoname}{\text{geometry}}

	
	\begin{abstract}
Although spatial indexes shorten the query response time, they rely on
complex tree structures to narrow down the search space. Such
structures in turn yield additional storage overhead and take a
toll on index maintenance. Recently, there have been a flurry of
efforts attempting to leverage Machine-Learning (ML) models to
simplify the index structures. However, existing geospatial
indexes can only index point data rather than complex geometries
such as polygons and trajectories that are widely available in
geospatial data. As a result, they cannot efficiently and
correctly answer geometry relationship queries.
This paper introduces {\indextitle}, an indexing mechanism for spatial
relationship queries on complex geometries. To achieve that,
{\indextitle} transforms geometries to Z-address intervals, and
then harnesses an existing order-preserving learned index to model
the cumulative distribution function between these intervals and
the record positions. The lightweight learned index greatly
reduces indexing overhead and provides faster or comparable query
latency. Most importantly, {\indextitle}
augments spatial query windows to support queries exactly for
common spatial relationships. Our experiments on real-world and
synthetic datasets show that {\indextitle} has
80\%-90\% lower storage overhead 
than Quad-Tree and 60\% - 80\% than R-tree  and 30\% - 70\% faster query
on medium selectivity. Moreover,
{\indextitle}'s maintenance throughput is 1.5 times higher on
insertion and 3 - 5 times higher on deletion.


\end{abstract}

	\maketitle
	
	\section{Introduction}


Database Management Systems (DBMSs) often create spatial indexes such as R-Tree~\cite{G84}, JED-Tree~\cite{B75}, and Quad-Tree~\cite{S84} to accelerate queries on geospatial data. Although spatial index structures shorten query response time, they rely on complex tree structures to narrow down the search space. Such structures in turn yield additional storage overhead and take a toll on index maintenance~\cite{YS17}. Recent works on spatial indices~\cite{OTH+17,TLB+21} mostly focus on accelerating query speed at the cost of even higher storage overhead. As depicted in Table~\ref{tab:storage-overhead}, R-Tree and Quad-Tree usually cost 10\% - 20\% additional storage overhead. This leads to significant dollar cost especially now that most enterprises move their data to cloud storage for better security and stability. Table~\ref{tab:cloud-cost} shows the cloud storage cost we collect from Amazon Web Services (AWS) EC2, the most popular cloud vendor. Such storage services are typically charged on a monthly or even an hourly basis, with additional fees for data transfer, which can result in unexpected bills for the users.

\begin{table}[h]
	\caption{Index storage overhead and storage dollar cost}	
	\vspace{-5mm}
	\begin{subtable}[h]{0.5\textwidth}
		\centering
		\caption{Index storage overhead on geospatial data (described in Table~\ref{tab:datasets})}
		\vspace{-2mm} {
		\footnotesize
		\begin{tabular}{|l|l|l|}
			\hline
			& Boost R-Tree & GEOS Quad-Tree \\ \hline\hline
			ROADS (8.3GB)         & 1.57GB        & 1.98GB         \\ \hline
			LinearWater (6.4GB) & 461.1 MB        & 685 MB          \\ \hline
			Parks (8.5GB)       & 783 MB        & 1.01GB   \\ \hline
		\end{tabular} }
	\label{tab:storage-overhead}
	\end{subtable}
	\begin{subtable}[h]{0.5\textwidth}
		\hfill
		\centering
		\caption{Cloud storage cost on Amazon Web Services EC2 instances}
		\vspace{-2mm}{
		\footnotesize
		\begin{tabular}{|l|l|l|}
			\hline
			RAM (no. CPU)      & SSD storage       & Data transfer          \\ \hline\hline
			32GB (2): \$0.16/hour  & L1: \$0.08/GB/month & Internal 1: \$0.01/GB \\ \hline
			64GB (4): \$0.33/hour  & L2: \$0.10/GB/month & Internal 2: \$0.02/GB  \\ \hline
			128GB  (8): \$0.66/hour & L3: \$0.12/GB/month & External: \$0.09/GB    \\ \hline
		\end{tabular}}
		\label{tab:cloud-cost}
	\end{subtable}
\end{table}

On the other hand, as open data lake formats
such as Apache Parquet~\cite{parquet} and GeoParquet~\cite{geoparquet}, become widely adopted, and as the trend to separate storage and computation layers in the cloud continues, 
researchers and practitioners are increasingly focusing their efforts on leveraging lightweight index structures. This strategy is aimed at enhancing data skipping efficiency at the storage level. Several works leverage
data synopses such as min-max, bounding box, and
histograms~\cite{SK+13,HKI+18,YS16,YS17} from the indexed data to
navigate queries. They adopt much simpler data
structure, which bring down the cost of storing and maintaining the
index. However, they compromise on query response time and cannot be
easily tailored to geospatial data (e.g., polygons, trajectories, etc.).



Recently, there have been a flurry of
works~\cite{KBC+18,WYT+19,DMY+20} attempting to leverage
Machine-Learning (ML) models to simplify the index structures. An
index, denoted as $y = f(x)$, can be viewed as an ML model, where x is
the lookup key and y is the physical position of the complete record
in an array.  Theoretically, this ML model learns the Cumulative
Distribution Function (CDF) between keys and their positions in a
sorted array.  Although learned index structures demonstrate promising
results on space saving and query speedup as opposed to the
traditional B+ Tree index, these approaches only work for
1-dimensional (1-D) sortable values

To remedy that, follow-up works extend the idea to support
geospatial points. These approaches~\cite{NDA+20,LLZ+20,WFX+19}
partition the multidimensional space to cells and assign IDs to these
cells using space-filling curve (e.g., Z-order curve~\cite{WFX+19}) or
mathematical equations~\cite{LLZ+20}. They can reduce data
dimension to 1-D and thus can be indexed by learned indexes. They
work well for geospatial points but are incapable of
handling complex geometries such as polygons and trajectories which
are widely available in geospatial data. One reason is that  complex
geometries can intersect multiple cells and thus have more than one 
IDs. This leads to duplicates in the final result~\cite{YZS19} and introduces additional challenges in index maintenance. In
addition, the user must hand-tune the partitioning resolution to find 
an appropriate cell size that is small enough but also does not
introduce too many duplicates. Finding such a sweet spot can be
prohibitively expensive especially when indexed geometries go across
large regions by nature (e.g., trajectories).

Designing a learned index structure for complex geometries presents
several major challenges, stated as follows: (1) \emph{Shapes.}
Geometries are collections of various complex shapes including
polygons and trajectories, which have been standardized to 7
categories~\cite{geometries}. Geometries are not 1-D or 2-D point values. Thus we cannot establish a CDF from such data to their positions for
an existing learned indexes. (2) \emph{Spatial distribution.} Geometries often show skewed spatial distributions in the space. For example, most landmarks, such as parks, hospitals, and government buildings, cluster at major metropolitan regions. The index structures should adapt to such distributions for better prediction performance. (3) \emph{Spatial relationship.} Geometries may have various spatial relationships such as $Contains$, $Intersects$, $Touches$, and $Disjoint$. Given a spatial query, the learned index structures must return all geometries that satisfy the spatial relationship to the query geometry.





This paper proposes {\indextitle}~\footnote{GLIN GitHub repository: https://github.com/DataOceanLab/GLIN}, a lightweight learned indexing mechanism for spatial range queries on complex geometries such as points, polygons, and trajectories. {\indextitle} by design produces low storage and maintenance overhead while achieving competitive query performance in common cases, as opposed to Quad-Tree and R-Tree. Moreover, {\indextitle} can work in conjunction with existing regular learned indexes to enable geospatial data support. Our contributions in this paper are summarized as follows:

$\bullet$ {\indextitle} transforms geometries to 1-D
sortable values using Z-order curve. We prove that {\indextitle} can always deliver correct results for both $Contains$ and $Intersects$ spatial relationships.

$\bullet$ For any existing order-preserving learned indexes
(Section~\ref{sec:zaddress}), {\indextitle} can extend it to index Z-address values and further improves the search performance by introducing additional information in leaf models. To the best of our knowledge, it is the first indexing mechanism that enables learned indexes on non-point data.

$\bullet$ {\indextitle} equips efficient algorithms to update its structure for data insertion and deletion while offering the
query accuracy guarantees.

$\bullet$ Our experimental analysis on real-world dataset shows that\
{\indextitle} has 80\% - 90\% lower storage overhead than Quad-Tree
and 60\% - 80\% than R-tree. Meanwhile, {\indextitle} has faster  query 
response time on  medium selectivity. Its update
throughput is 1.5 times higher on insertion and 3 ~ 5 times
higher on deletion.

	\section{Background}
\label{sec:background}

{\bf Spatial range query.} Given a query window $Q$, a spatial dataset $R$ and a predefined spatial relationship $SR$, a range query denoted as $range(Q, R, SR)$ finds the geometries in $R$ such that each geometry (denoted as $GM$) has $SR$ relationship with $Q$. $GM$ and $Q$ can have any shapes including polygons and lines.

{\bf Spatial relationship.}
SQL/MM3 standard~\cite{geometries} lists a number of possible spatial
relationships between two geometries. This includes but is not limited
to: Contains, Intersects, Touches, and Disjoint. In this paper, we
focus on the two most common spatial relationships: (1) $Contains$
(Figure~\ref{fig:zorder} Case 1): given two geometries $Q$ and $GM$,
"Q contains GM" is true if and only if no points of GM lie in the
exterior of Q, and at least one point of the interior of GM lies in
the interior of Q. (2) $Intersects$ (Figure~\ref{fig:zorder} Case 1,
2, 3): given two geometries $Q$ and $GM$, "Q intersects GM" is true if
Q and GM share any portion of space. $Contains$ is a special case of
$Intersects$. If "Q contains GM" is true, then "Q intersects GM" must
be true as well.

{\bf Minimum Bounding Rectangle (MBR).}
An MBR describes the maximum extents of a 2-dimensional geometry in an
$(x, y)$ coordinate system. An MBR consists of four values, the minimum
and maximum values of $x$ and $y$ coordinates of a geometry, and are
represented as two points, $p_{min}(x_{min}, y_{min})$ and
$p_{max}(x_{max}, y_{max})$ (see Figure~\ref{fig:zorder} Case 1). The
coordinates of the MBR can be easily obtained by iterating every
coordinate of a geometry. MBR is often used to approximate
geometries since it is a much simpler shape.

{\bf Probing and Refinement steps of spatial index search.}
Most existing spatial indexing mechanisms, such as R-Tree, Quad-Tree, and KD-Tree, approximate complex geometries to their MBR and then build index structures on MBRs. A spatial range query is processed mostly in two steps. (1) $Index~probing$: the MBR of the query window is processed against the spatial index. It returns a set of candidate geometries whose MBRs possibly $intersect$ the query window MBR. This set of candidates is not the exact answer of this query but is a super set of the answer. (2) $Refinement$: the candidate geometries are checked against the query window using their actual shapes with a spatial relationship such as $Contains$ or $Intersects$. The refinement step is computationally expensive due to the complexity of shapes and usually takes more time than the probing step~\cite{BM19}. {\indextitle} also follows this two-step process which can significantly reduce computation cost and index storage overhead. Existing learned spatial indexes~\cite{NDA+20,LLZ+20,WFX+19} only perform the probing step and their results might only be a subset of the exact answer if the underlying data is not points.


%
%

	\section{Overview}
\label{sec:overview}

\begin{figure}
	\includegraphics[width=0.9\linewidth]{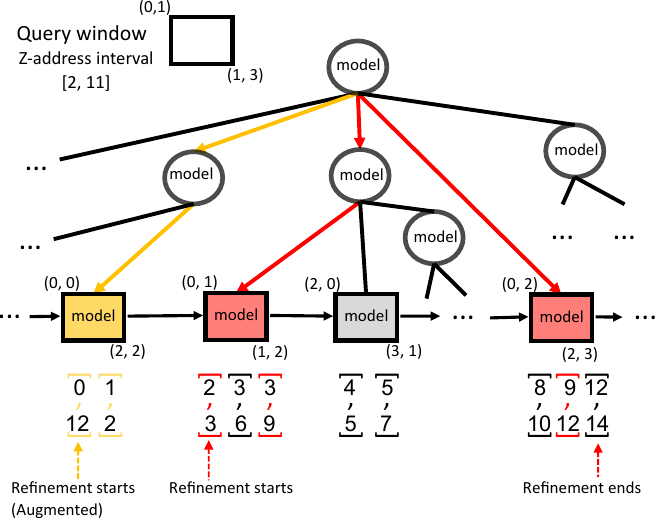}
	\caption{{\indextitle} index structure. White nodes are internal nodes and colored nodes are leaves. Index search: (1) $Contains$: follow the red paths and return red records. The gray node is skipped as its MBR does not intersect the query. (2) $Intersects$: follow the yellow path and the second red path. Red and yellow records will be returned.}
	\label{fig:overview}
	\vspace{-5pt}
\end{figure}

The index structure of {\indextitle} is depicted in Figure~\ref{fig:overview}.

\textbf{Z-address interval.}
To establish this CDF and enable the model training process, {\indextitle} assigns each geometry a Z-address interval, an one-dimensional sortable interval (serve as keys), by using a well-known space-filling curve called Z-order curve. In this paper, we also study that how $Contains$ and $Intersects$ relationships are reflected on Z-address intervals and prove that {\indextitle} can guarantee the query accuracy in both cases.

\textbf{Index structure.}
Once the CDF is made available between Z-address intervals and record positions, {\indextitle} can harness an existing order-preserving learned indexes, called the base index, such as ALEX~\cite{DMY+20} and RadixSpline~\cite{KMR+20} to model the CDF. Since Z-address intervals cannot 100\% preserve the original shape information and spatial proximity, the index probing will return some false positive results. In response, {\indextitle} introduces a refinement phase to prune out all false positives. In addition, it creates a MBR on each leaf node of the hierarchical model to accelerate the refinement phase.


\textbf{Query augmentation.}
The basic indexing mechanism in {\indextitle} is designed to handle $Contains$ relationship and may produce true negatives for $Intersects$ relationship. To remedy that, {\indextitle} employs a piecewise function with outlier handling to augment the query window. More precisely, {\indextitle} will enlarge the Z-address interval of the query window to make sure that it covers all correct results at the cost of additional pruning time.

	\section{Z-address intervals and order preserving indexes}
\label{sec:zaddress}

\begin{table}
	\small
	\caption{Notations used in this paper}	
	\resizebox{\linewidth}{!}{
	\begin{tabular}{|p{0.1\linewidth}|p{0.9\linewidth}|}
		\hline
		\textbf{Term}                   & \textbf{Definition}           \\ \hline
		Q, GM                  & Q - a spatial range query window. GM - an indexed geometry. Both can be in any shapes.\\ \hline
		MBR       & Minimum Bounding Rectangle of a geometry, represented as two points $p_{min}(x_{min}, y_{min})$ and $p_{max}(x_{max}, y_{max})$  \\ 
		       & $MBR_Q$ - MBR of Q.  $MBR_{GM}$ - MBR of GM. \\ \hline		
		Zmin & Z-address for $p_{min}$ of a MBR                \\
		 & $Zmin_Q$ - Zmin of Q, $Zmin_{GM}$ - Zmin of GM.                  \\ \hline		
		Zmax    & Z-address for $p_{max}$ of a MBR                   \\
		 & $Zmax_Q$ - Zmax of Q, $Zmax_{GM}$ - Zmax of GM.                  \\ \hline		
		Zitvl & Z-address interval described by $\itvl{Zmin, Zmax}$                 \\
		 & $Zitvl_Q$ - Zitvl of Q, $Zitvl_{GM}$ - Zitvl of GM.                  \\ \hline				
	\end{tabular}
}
\vspace{-10pt}
\label{tab:notations}
\end{table}

Notations used in this section are summarized in Table~\ref{tab:notations}.

\begin{figure}
	\includegraphics[width=0.9\linewidth]{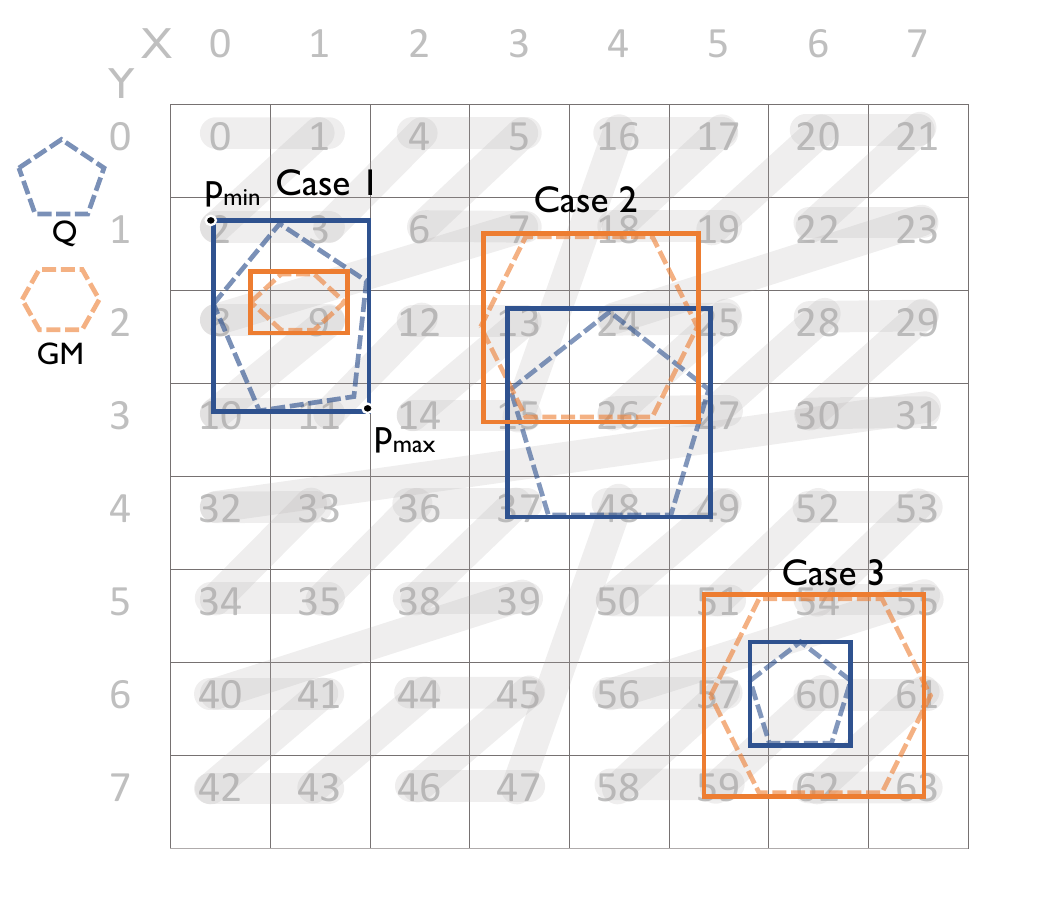}
	\vspace{-10pt}	
	\caption{Spatial relationship. Case 1: Q $contains$ GM; Case 1, 2,
	3: Q $intersects$ GM. Z-address($p_{min}$) = 2,
	Z-address($p_{max}$) = 11, Z-address interval of Q in Case 1 =
	$\itvl{2, 11}$.}
	\label{fig:zorder}
	\vspace{-5pt}
\end{figure}

{\bf Z-address.}
Z-order curve is a Z-shape curve (see Figure~\ref{fig:zorder}) that connects all
2-dimensional positive integer coordinates in the space. Each
coordinate $(x, y)$ will then have a Z-address. The Z-addresses of two
nearby coordinates will likely be close to each other. For example,
$p_{min}(0, 1)$ in Figure~\ref{fig:zorder} will have a Z-address 2.
{\indextitle} rounds down geospatial coordinates to their nearest
integers: x = $\frac{longitude-(-180^{\circ})}{cell~size}$ and y =
$\frac{latitude-(-90^{\circ})}{cell~size}$. Then it calculates the
Z-address using libmorton~\cite{libmorton} which interleaves the
binary representation of x and y coordinates~\cite{LZL+07}.  GLIN sets the cell size as $5\times 10^{-7}$ to represent centimeter-level
precision~\cite{degree-precision}. Detailed discussion can be found in Section~\ref{subsec:cellsize}.


{\bf Z-address interval.}
{\indextitle} assigns a Z-address interval, Zitvl $\itvl{Zmin, Zmax}$,
for every geometry including the indexed geometries and the range
query window. It is computed in two steps: (1) find the
MBR of a geometry, represented as $p_{min}(x_{min}, y_{min})$ and
$p_{max}(x_{max}, y_{max})$; (2) compute the minimum and maximum
Z-addresses $Zmin, Zmax$ from $p_{min}, p_{max}$ respectively. In
Figure~\ref{fig:zorder} Case 1, Q has $Zitvl_Q \itvl{2,11}$.

{\bf Notes.}
(1) When calculating $Zitvl$, {\indextitle} must use $p_{min}$ and $p_{max}$ rather than vertices of the geometry because $Zitvl$ from the latter might not cover all Z-addresses touched by the geometry. For example, vertices of Q in Figure~\ref{fig:zorder} Case 1 indicate $Zitvl$ = 
$\itvl{3,11}$, which misses Z-address 2. (2) We choose Z-order curve due to its monotonic ordering property~\cite{LZL+07} which guarantees that $Zitvl$ from $p_{min}$ and $p_{max}$ covers the Z-address of any point that falls inside this geometry~\cite{WFX+19}. In Hilbert curve (or other space filling curves), the $Hmin$ and $Hmax$ of the desired $Hitvl$ are actually on the boundary of the MBR (details omitted due to page limit). To obtain such $Hitvl$, we have to calculate every H-address touched by the MBR. This will significantly slow down the queries and require to tune the cell size which cannot be too large or too small.

{\bf Spatial relationship between intervals.}
Since {\indextitle} is built on Z-address intervals, when we examine
the spatial relationship between Q and GM, we also need to understand
how this relationship translates to Z-address intervals of $Q$ and
$GM$, denoted as $Zitvl_Q$ and $Zitvl_{GM}$, respectively.  We show
two important lemmas below which allow us to prune the search space
for $Contains$ and $Intersects$ relationship using range scans.  The
proofs are given in Appendix~\ref{sec:proofs} in the interest of
space.

\begin{lem}\label{lem:contains}
	{\bf Z-address interval \underline{Contains}.} If Q $contains$ GM, then $Zitvl_Q~contains~Zitvl_{GM}$
\end{lem}

where $Zitvl_Q~contains~Zitvl_{GM} \iff  Zmin_Q \leq Zmin_{GM} \leq Zmax_Q \land Zmin_Q \leq Zmax_{GM} \leq Zmax_Q$.

\begin{lem}\label{lem:intersects}
	{\bf Z-address interval \underline{Intersects}.} If Q $Intersects$ GM, then $Zitvl_Q~intersects~Zitvl_{GM}$
\end{lem}

where $Zitvl_Q~intersects~Zitvl_{GM} \iff 
Zmin_Q \leq Zmax_{GM} \wedge Zmax_Q \geq Zmax_{GM}
$. 
In other words, $Zitvl_Q$ and $Zitvl_{GM}$
share some portion of the intervals

\textbf{Order-preserving learned index.}
GLIN is designed as a general framework to adapt 1-D learned indexes
as spatial indexes for polygon relationship queries. One important
property that the underlying index must satisfy is
\emph{order-preserving}. To
explain why, we first define the order-preserving property.

\begin{definition}
Let $I$ be a 1-D range index where the leaves store item keys. The
keys in the leaves can be conceptually concatenated into a list in
the pre-order traversal of the index.  We say a 1-D range index is
\emph{order-preserving} if the list is in sorted order.
\label{def:order_preserving}
\end{definition}

Order-preserving is crucial for us to correctly prune the search
search space in tree probing. Suppose we index the geometries with an
order-preserving range index using $Zmin$ of the geometries.
As we will show later, Lemma~\ref{lem:contains}
and~\ref{lem:intersects} allow us to search for geometrices with
respect to a query geometry $Q$ using a sufficiently large $Zitvl$.
Taking $Contains$ as an example, the range can be $\itvl{Zmin_Q, Zmax_Q}$.
Since the index is order preserving, we can simply probe the index for
the first geometry with $Zmin_{GM} \geq Zmin_Q$ and scan the leaf
levels until $Zmin_{GM} > Zmax_Q$. However, we cannot do so if an
index is not order preserving as there might be some $Zmin_{GM} >
Zmin_Q$ that appears before the first $Zmin_Q$ (if it exists) in the
index.

In traditional range indexes such as B-tree (which technically can also
be used in GLIN), the list consists of all keys in the leaf level from
left to right. It is order-preserving because the tree strictly
divides sub-trees into disjoint and increasing key ranges from left to
right. Many learned indexes are also order-preserving. For instance,
ALEX~\cite{DMY+20} also divides its key into disjoint and increasing
key ranges based on linear regression models, and thus it can be used
in GLIN.  RMI~\cite{KBC+18} is an example of non-order-preserving
index because an implementation can choose not to enforce the
monotonicity of assignment of keys in the internal models when they
are complex neural networks. Consequently, we cannot use RMI (or adapt
any RMI based spatial indexes such as RSMI~\cite{QLJ+20} to support
exact spatial relationship queries).


	\section{Index initialization}
\label{sec:initialize}
 
To create an index, {\indextitle} reads a set of geometry records and initializes the index structure based on the geometries. 
The mechanism depicted in this section only handles the spatial range query with the $Contains$ spatial relationship, which is "the query window contains geometries".

\textbf{Sort geometries by Z-address intervals.}
To establish the CDF between keys and record positions, the first step is to put the geometries in a sequential order. 
{\indextitle} sorts geometries based on their Z-address intervals (see Section~\ref{sec:zaddress}). The reason is two-fold: (1) Z-addresses can partially preserve the spatial proximity of geometries. 
This is critical because a spatial range query looks for geometries that lie in the same region. This query will inevitably scan a large portion of the table if the trained ML model does not respect any spatial proximity. (2) A Z-address interval can partially preserve the shapes of geometries. This will make it possible for {\indextitle} to employ different strategies for $Contains$ and $Intersects$ relationships for the sake of query performance.

{\indextitle} iterates through every geometry and calculates the Z-address interval (i.e., $Zitvl = \itvl{Zmin, Zmax}$) for this geometry. {\indextitle} then sorts these geometries by the $Zmin$ of their intervals. In the rest of the index initialization phase, $Zmax$ of intervals will be completely discarded. In other words, {\indextitle} actually indexes $<Zmin, geometry~key>$ pairs while traditional spatial indices index $<MBR, geometry~key>$ pairs. Later, in the index probing phase of the index search, {\indextitle} only checks if the Z-address interval of the query window contains $Zmin$ of geometries (i.e., $Zmin_Q \leq Zmin_{GM} \leq Zmax_Q$). According to Lemma~\ref{lem:contains}, this introduces some false positive results but does not have true-negatives. $Zmax$ will be used in Section~\ref{sec:piecewise} for query augmentation.
 \begin{figure}
	\centering
	\begin{minipage}{.15\textwidth}
		\includegraphics[width=\linewidth]{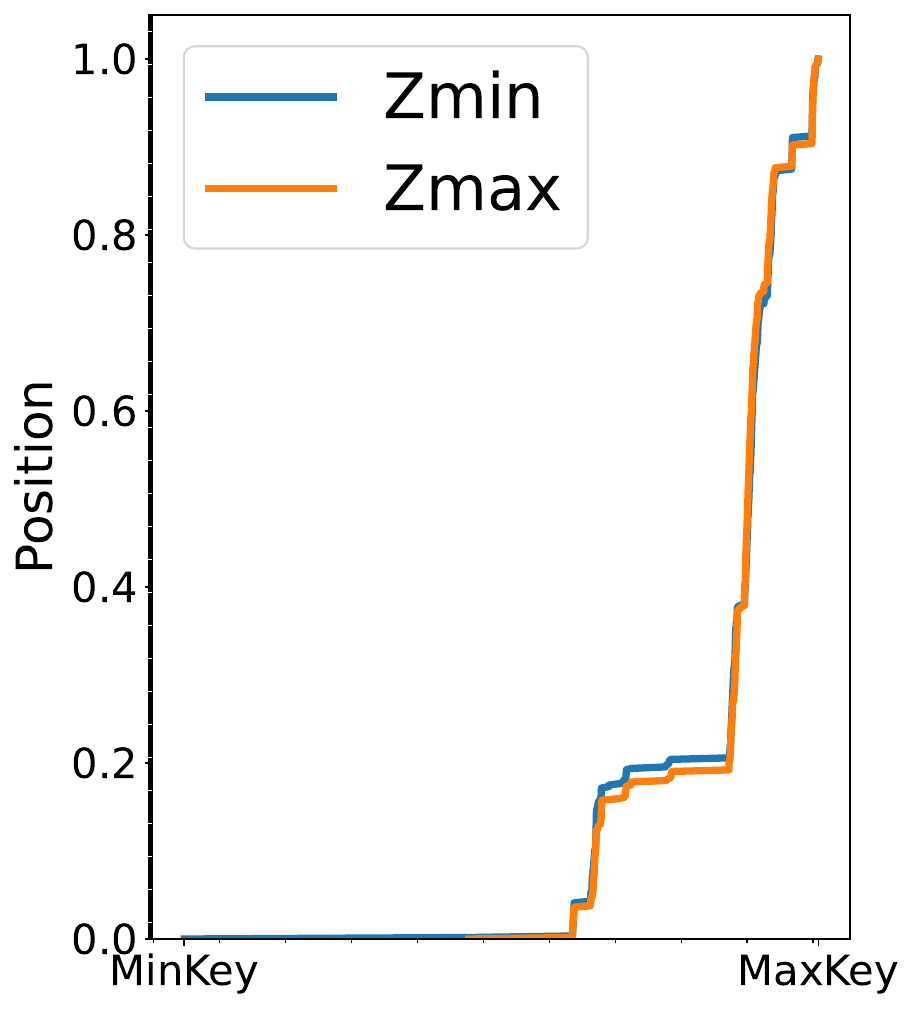}
		\subcaption{LW}
	\end{minipage}
	\begin{minipage}{.15\textwidth}
		\includegraphics[width=\linewidth]{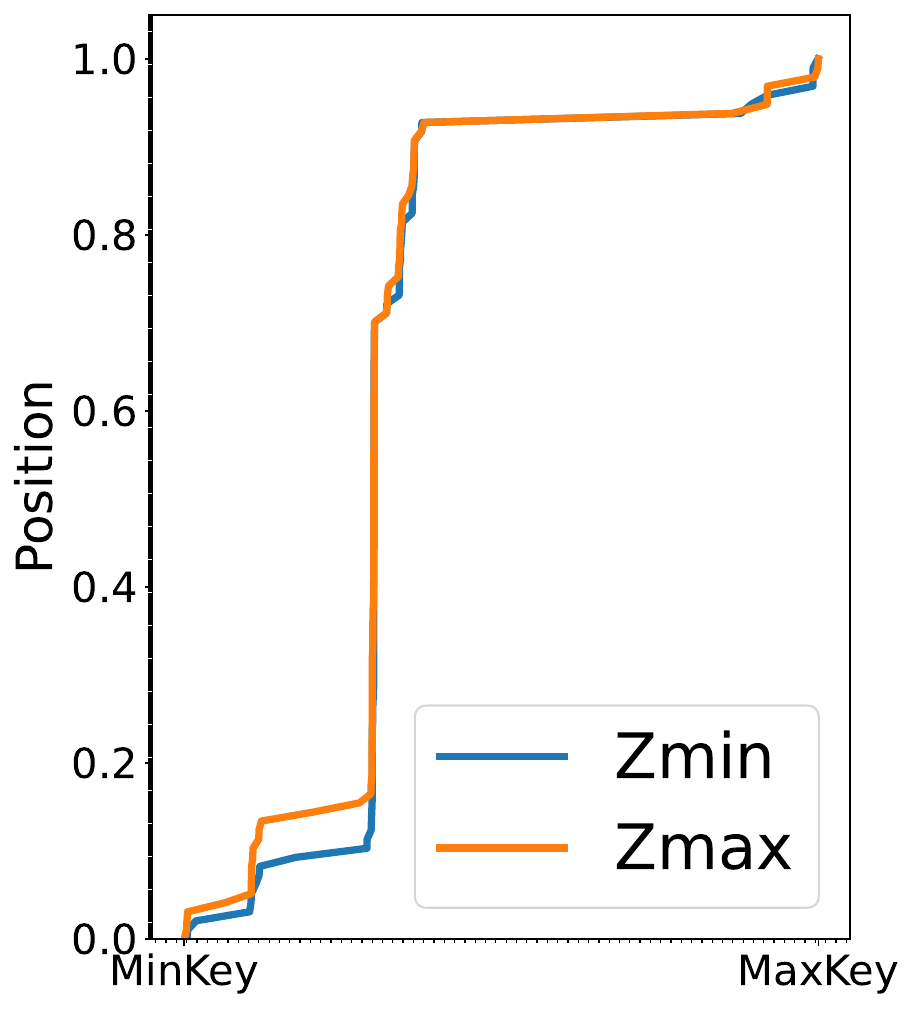}
		\subcaption{PARKS}
	\end{minipage}
	\begin{minipage}{.15\textwidth}
		\includegraphics[width=\linewidth]{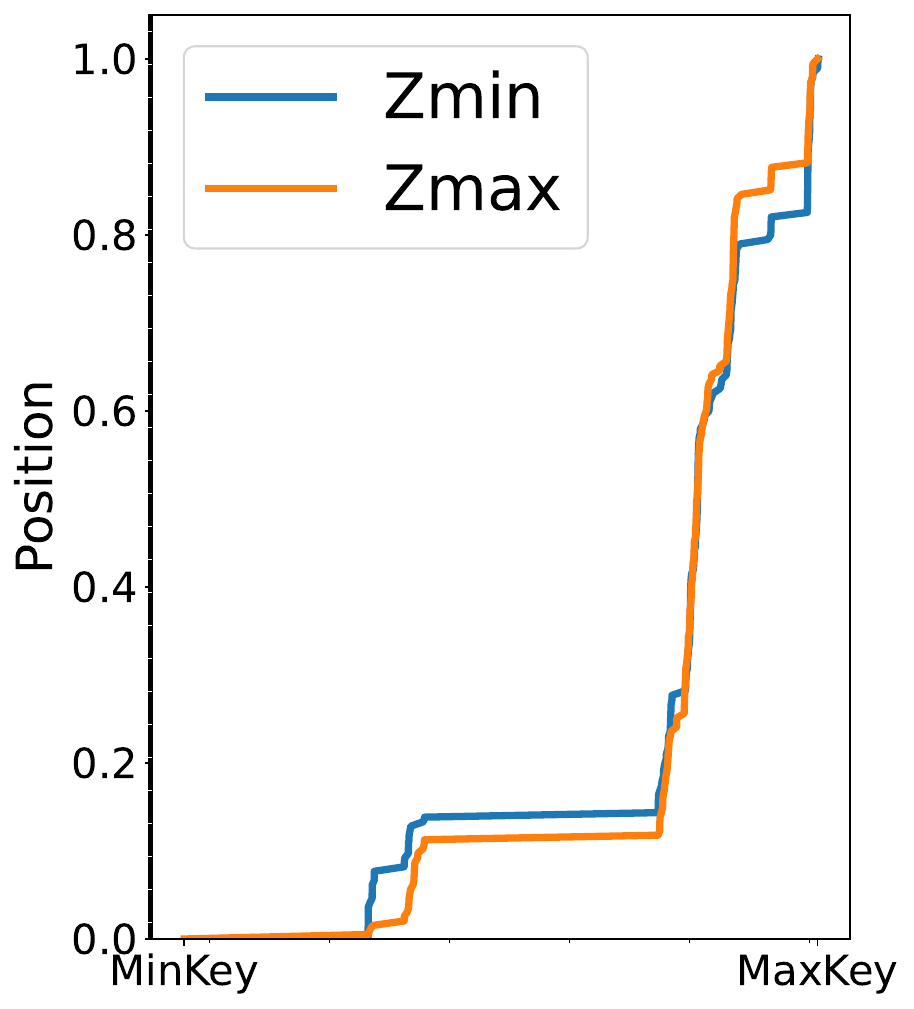}
		\subcaption{ROADS}
	\end{minipage}
	\caption{CDFs of different datasets based on $Zmin$ and $Zmax$}
	\label{fig:datasetcdf}
	\vspace{-10pt}
\end{figure}
It is worth noting that {\indextitle} could also sort geometries by the $Zmax$ instead of $Zmin$. However, this will not make much difference in the learned CDF models since $Zmax$ will follow the same data distribution of $Zmin$ (see Figure~\ref{fig:datasetcdf}). In Section~\ref{sec:piecewise}, both $Zmin$ and $Zmax$ will be needed in order to handle $Intersects$ relationship.

\textbf{Build the base learned index.} Once geometries are sorted by the $Zmin$ of their intervals, {\indextitle} will train a base index to learn the CDF between these $Zmin$ addresses and the record positions. {\indextitle} works in conjunction with any order-preserving learned index and extend it to uphold geometries. These indexes usually possess a hierarchical structure to build models for different regions. In this paper, we adopt the method in ALEX~\cite{DMY+20} because it supports index updates by design.

\textbf{Create MBRs in leaf nodes.} 
Since {\indextitle} trains and queries models based on Z-addresses of geometries, the model prediction may introduce more false positives that need to be pruned during the refinement step.

To mitigate this, {\indextitle} employs a simple yet efficient method to reduce the computation cost. When constructing each leaf node of the hierarchical model, {\indextitle} also creates a MBR of all geometries in this node. This can be done by traversing all geometries and finding the overall $p_{min}(x_{min}, y_{min})$ and $p_{max}(x_{max}, y_{max})$. When refining the query results, {\indextitle} will directly skip a leaf node unless the MBR of this node intersects the query window's MBR.

\textbf{Index maintenance.} 
For insertion, {\indextitle} takes as input a geometry key and inserts it to the index. To insert a new record, {\indextitle} first obtains the $Zmin$ address for the geometry key in this record using the approach described in Section~\ref{sec:zaddress}, then inserts the record to the base index.  Once {\indextitle} puts the new record in a leaf node, it expands the MBR of the leaf node to include the new geometry.

For deletion, the input is a geometry key and {\indextitle} deletes records that have the same key. Similar to the insertion, the first step to delete a geometry key is to get the $Zmin$ address of this geometry. It is possible that several different geometries share the same  $Zmin$ addresses. In that case, {\indextitle} only erases records that have the same geometry key. The MBR of the involved leaf node will not be shrunk after the deletion because it requires a scan of all records in this leaf node to get the latest MBR. However, this does not affect the correctness of {\indextitle} as the out-of-date MBR only introduces more false positives instead of true negatives.
	\section{Index Search}\label{sec:search}
\begin{algorithm}
	\footnotesize
	\SetKwInOut{Input}{Input}\SetKwInOut{Output}{Output}
	\Input{A query window $Q$, a spatial relationship $SR$}
	\Output{A set of records $Result$ that satisfy $SR$ with $Q$}
	\tcc{Step 1: index probing}	
	$Zitvl_Q\itvl{Zmin, Zmax}$ = calculate\_zitvl($Q$);\\
	\If{SR is "$Intersects$" relationship}
	{
		\tcp{Augment the query window}
		$Zmin$ = augment($Zitvl_Q$, piecewise function).$Zmin$;\\ 
	}
	start\_position = $model\_traversal({\indextitle}.root, Zmin)$;\\
	\tcc{Step 2: refinement}
	$Result$ = new List();\\
	iterator = start\_position;\\
	\While{iterator.key $\leq Zmax$}
	{
		$geom$ = Get\_Record(iterator);\\
		\If{$Q$ has $SR$ with $geom$}
		{
			$Result$.add($geom$);\\
		}
		node = iterator.node();\\
		\eIf{iterator $\geq$ node.last\_position()}
		{
			\While{$MBR_Q$ does not intersect $node.MBR$}
			{
				node = node.next\_node();\tcp{Skip this node}
			}
			iterator = node.first\_position();\\
		}
		{iterator++;\\}
	}
	\textbf{Return} $Result$;\\
	\SetKwFunction{FMain}{model\_traversal (in the base index)}
	\SetKwProg{Fn}{Function}{:}{}
	\caption{{\indextitle} Index Search}
	\label{algo:search}
\end{algorithm}


\subsection{Probe the base index}

{\indextitle} leverages an existing learned index to build the hierarchical model based on the $Zmin$ addresses of geometries. Therefore, to search the index, this algorithm must first obtain the Z-address interval of the query window (see Section~\ref{sec:zaddress}). The Z-address interval $\itvl{$Zmin$, $Zmax$}$ of the query window will then serve as the actual input for the index probing which finds geometries whose $Zmin$ is within the interval. According to Lemma~\ref{lem:contains}, geometries whose $Zmin$ is not within this interval are guaranteed not to be contained by the query window.


The process of searching the hierarchy model is identical to the base learned index on which {\indextitle} is built.  As described in Algorithm~\ref{algo:search}, the index probing step requires a tree-like traversal of the hierarchical model (i.e., $model\_traversal$). It uses $Zmin$ address of the query window as the lookup key (see the first red pathin Figure~\ref{fig:overview}). This traversal runs in a top-down fashion starting from the root. The model inside the root node will predict a position in the pointer array using the lookup key. Once the algorithm reaches the leaf node, it will first perform the model prediction to find an approximate position in the record array and then run an exponential search from this position to locate the correct result.

\subsection{Refine the results}

The records returned by the index probing have some false positives due to the following reasons: (1) {\indextitle} uses the MBR of each geometry to produce the Z-addresses rather than the actual shape. (2) The Z address interval includes additional Z-addresses. For example, in Case 1 of Figure~\ref{fig:zorder}, the $MBR_Q$ only contains 6 addresses (2, 3, 8, 9, 10, 11) but all records whose $Zmin$ values are in$ \itvl{2, 11}$ (the Zitvl of Q, 10 Z-addresses in total) will be returned by the index probing step. (3) the hierarchical model is built upon the  $Zmin$ of geometries without considering $Zmax$ at all.

As given in Algorithm~\ref{algo:search}, {\indextitle} conducts a refinement step to filter out these false positive results and hence offer accuracy guarantee. As illustrated in Figure~\ref{fig:overview}, this refinement starts from the position returned by the $Zmin$-based model traversal and keeps checking if every geometry satisfies the spatial relationship with the query window using their exact shapes, until it arrives at the position whose key is no larger than the query window's $Zmax$. The MBRs in leaf nodes will help {\indextitle} skip non-intersecting nodes directly.

\begin{table}[]
	\scriptsize
	\centering
	\caption{Number of records checked during the refinement}	
	\begin{tabular}{|r|r|r|r|}
		\hline
		& Query selectivity & W/o leaf MBR & W/ leaf MBR \\ \hline\hline
		\multirow{3}{*}{ROADS} & 1\%         & 3339990      & 369184      \\ \cline{2-4} 
		& 0.1\%       & 1173710      & 67474     \\ \cline{2-4} 
		& 0.01\%      & 632839       & 18244        \\ \hline
		\multirow{3}{*}{PARKS} & 1\%         & 1126520      & 154685     \\ \cline{2-4} 
		& 0.1\%       & 291197       & 21700    \\ \cline{2-4} 
		& 0.01\%      & 105076       & 4180    \\ \hline
	\end{tabular}
	\label{tab:refinement_mbr}
	\vspace{-10pt}
\end{table}

\section{Query augmentation for Intersects}\label{sec:piecewise}

\subsection{The lowest intersecting Z-address}

Given a query window $Q$, Algorithm~\ref{algo:search} in Section~\ref{sec:search} finds
all geometries $GM$ such that $Zmin_{GM} \geq Zmin_Q$. However, an
intersecting geometry $GM$ may have $Zmin_{GM} < Zmin_Q$ and
$Zmax_{GM} \geq Zmin_Q$ (Lemma~\ref{lem:intersects}). Therefore, the
algorithm does not return a superset of all the intersecting
geometries. For example, Case 2 in Figure~\ref{fig:zorder} shows two
intersecting polygons whose Z-address intervals are $\itvl{7, 27}$ and
$\itvl{13, 49}$. Let the first one be an indexed geometry $GM$ and the
second one be the query window $Q$. $\itvl{7, 27}$ will be missing from the final results returned by Algorithm~\ref{algo:search}.

A simple solution is, for all
geometries that have $Zmax_{GM} \geq Zmin_Q$, we find the smallest $Zmin_{GM}$, namely the {\it
lowest intersecting Z-address}. Then we augment the query window by taking $min(Zmin_Q, lowest~intersecting~Z-address)$. Unfortunately, since the underlying records are sorted by their $Zmin_{GM}$ instead of $Zmax_{GM}$, finding such value for each query requires a full scan, which is prohibitively expensive. Hence, we need a data structure to help {\indextitle} quickly find $lowest~intersecting~Z-address$.

\subsection{The piecewise function}

The intuition is that we divide the domain range of $Zmax$ to a few sub-domains, and precompute the lowest intersecting Z-address for each sub-domain. Consider a set of $N$ Z-address intervals, let $Z$ be the maximum $Zmax$ among
all intervals. Note that all Z-addresses are non-negative and thus the $Zmax$ of all
intervals $\in$ $\itvl{0, Z}$. If we divide the range
$[0, Z]$ into $k$ disjoint domains: $\itvl{0, Z_1}$, $\itvllo{Z_1, Z_2}, \ldots
\itvllo{Z_{k-1}, Z}$,
we can define a
piecewise-constant function with these $k$ pieces and compute the
lowest intersecting Z-address for each. To simplify notation, we
denote $Z_0 = -1$ and $Z_k = Z$, and thus the $i^{th}$ piece can be
denoted as $\itvllo{Z_i, Z_{i+1}}$ for $0 \leq i < k$. Given a query
window $Q$,
if $Zmin_Q \in \itvllo{Z_i, Z_{i+1}}$, we can find such value on the
fly by binary searching the piecewise function, and then augment the
query window.  It is worth noting that the lowest intersecting
Z-address
monotonically increases over $i$\eat{of a
sub-range must be the smallest value for all subsequent sub-ranges}.
A proof sketch is shown in Appendix~\ref{sec:proofs}.

\begin{figure}
	\includegraphics[width=0.5\linewidth]{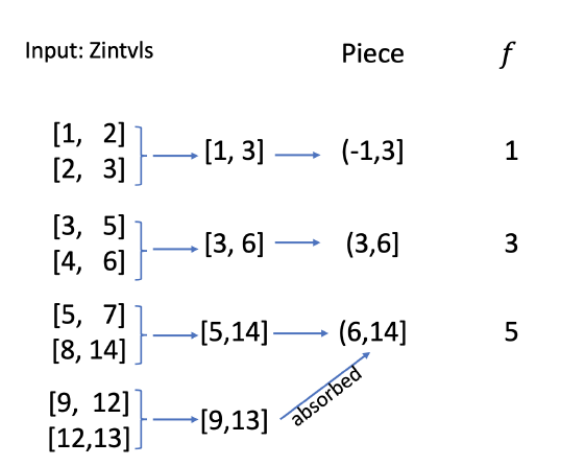}
	\caption{A piecewise funtion with piece granularity $m = 2$.}
	\label{fig:piecelimitation}
\end{figure}

Below is a possible piecewise function for Z-intervals listed in Figure~\ref{fig:piecelimitation}. If we want to search for geometries that intersect a query
window $Q$, whose z-address interval is $\itvl{2, 5}$, then we can use $f(2) =
1$ as the new $Zmin_Q$ to search the base index. Without this
function, we will miss  a potentially intersecting geometry $\itvl{1, 2}$ during the index probing phase.

\begin{align}\nonumber
\small	
	f(Zmin_{Q}) = 
	\begin{cases} 
		1 & -1 < Zmin_{Q} \leq 3 \\
		3 &  3 < Zmin_{Q} \leq  6 \\
		5 &  6 <  Zmin_{Q} \leq 14 \\
	\end{cases}
\end{align}\label{eqt:piecewise}
\begin{figure}
		\includegraphics[width= 0.6\linewidth]{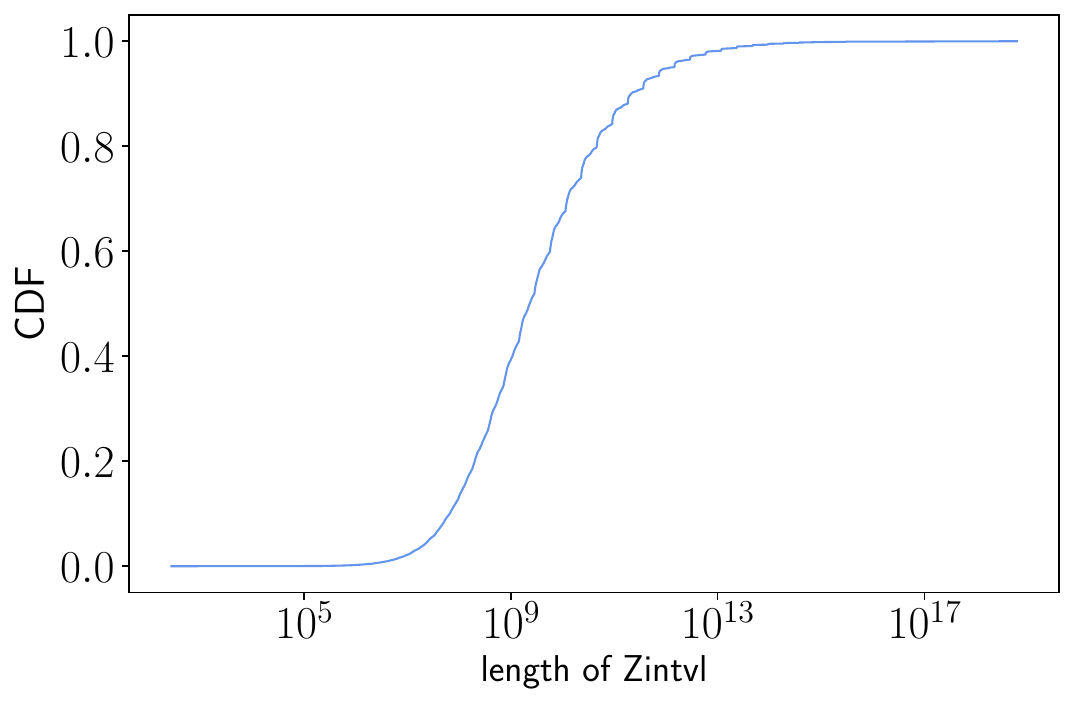}
		\vspace{0pt}
		\caption{CDF base on $Zintvl$ length}
		\label{fig:outlier}
		\vspace{-3mm}
\end{figure}

{\bf Issues with long intervals.} Long Z-address intervals
could jeopardize the effectiveness of the piecewise function. If we
insert interval $\itvl{0, 14}$ in the dataset, then the piecewise function
becomes a contant function with value $0$ everywhere. The reason is
that the new interval has to be considered in each lower bound of the
lowest intersecting z-address because $Zmax_{GM} = 14$ is greater than
the lower ends of all intervals. Fortunately, we observe that such
long intervals rarely appear in real-world datasets (Figure ~\ref{fig:outlier}). If we
treat them as outliers and separately index them in a different
structure, the piecewise function will still produce close lower
bounds. In GLIN, we first find out the 99\% percentile (an adjustable
threshold) of the lengths of the Z-address intervals as the outlier
length threshold. We treat all intervals with length longer than that
as outliers. They are separately maintained in a smaller outlier index
using the base index. For any intersects query, we additionally probe and search
the index from $Zmin_Q = 0$ using Algorithm~\ref{algo:search}.

{\bf Greedy construction of the piece-wise function.} There is a
trade-off of how many intervals we partition the range $\itvl{0,Z}$
into.  On one hand, we can create one interval per distinct $Zmax$
value in
the dataset, which provides the exact lowest intersecting z-address
for any query window. However, it takes $O(\log N)$ time to augment
the query window. On the other hand, we can create one single interval
$\itvl{0, Z}$, which maps to the smallest Zmin for all possible query
windows. Augmenting the query is quick with $O(1)$ time, but the
number of geometries to go through the refinement step will be very
large. Hence, we balance the trade-off by constructing the piecewise
function using the following greedy algorithm
(Algorithm~\ref{algo:piecewise_construction}): we sort all geometries'
intervals by $Zmin_{GM}$ (which can be done using a
leaf level traversal in
GLIN without an additional sorting operation),
and combine every $m$ intervals (called \emph{piece granularity}) into
a single combined interval that exactly covers the $m$ intervals. We
treat all combined intervals as the input dataset, denoted as
$\itvl{Zmin_1', Zmax_1'},$
$\ldots \itvl{Zmin_{\lceil N/m\rceil}', Zmax_{\lceil N/m \rceil}'}\}$.
Then we scan these intervals and create a new piece if the $Zmax$ of
an interval is greater than that of the previous one.
The higher $m$ is, the less accurate the
lower bounds are. A less accurate lower bound leads to more records to refine.

Figure~\ref{fig:piecelimitation} shows a concrete example of how to
build the piecewise function using
Algorithm~\ref{algo:piecewise_construction} with piece granularity of
2. The input $Zintvl$ is first
sorted by $Zmin_{GM}$ (and ties are broken using $Zmax_{GM}$) and we
combine every two intervals into a larger interval. For example, the
first two $\itvl{1, 2}$ and $\itvl{2, 3}$ are combined into $\itvl{1,
3}$.
Then we construct a new piece starting from the previous $Zmax = -1$
(exclusive) until the $Zmax = 3$ (inclusive) of the combined interval,
and record the function value $f = Zmin = 1$.
Note that, if the $Zmax$ of the combined interval is not larger than
the previous $Zmax$ value (e.g, the last combined interval $\itvl{9,
13}$), it should be absorbed by the previous pieces without any update
due to the monotonicity of the piecewise function.

\begin{algorithm}
	\footnotesize
	\SetKwInOut{Input}{Input}\SetKwInOut{Output}{Output}
	\Input{$Zintvls$ sorted by $Zmin$  $I$,  $Piece\_granularity$ $m$}
	\Output{ piecewise function $PW$}
	c = 0;\\
	prev\_zmax = -1; \\
	\ForEach{$Zintvl$ in $I$}
	{
		\eIf{c == 0}
		{current\_zmin = $Zintvl.Zmin$;\\ 
		current\_zmax = $Zintvl.Zmax$;}
		{ current\_zmax =max($Zintvl.Zmax$, current\_zmax) ;}
		c ++ ; \\
		\If{c == $m$}{
			\If {current\_zmax > prev\_zmax}{$PW$.pushback(current\_zmax, current\_zmin);\\
			prev\_zmax = current\_zmax; }
			counter = 0 ;\\
			}
	 }
	\If{c> 0 }{\If{current\_zmax > prev\_zmax}{$PW$.pushback(current\_zmax, current\_zmin);\\}}
		\textbf{Return} $PW$ ;\\
	\caption{Initialize the piecewise Function}
	\label{algo:piecewise_construction}
\end{algorithm}

{\bf Updating piecewise function.} To handle an update,
we may also need to update the piecewise functions. For insertion, suppose
the inserted geometry is $GM$. We first check its z-address interval
length against the outlier length threshold. If the length is greater
than the threshold, we do not need to update the piecewise function.
Otherwise, we use binary search to find the first interval $(Z_i,
Z_{i+1}]$ such that $Zmin_{GM}$ is in that interval. For all intervals
$j \geq i$, we update the function value of range $j$ to
$Zmin_{GM}$ if $Zmax_{GM} > Zmin_j$ with $\min\{Zmin_{GM}, f(Zmax_j)\}$. For example, 
if we insert a non-outlier geometry with its z-addres interval being
$\itvl{6,8}$, we first find the first piece that
contains its $Zmin_{GM}$ = 6 and scan forward until it no longer
overlaps with the piece. In this case, both the $\itvllo{3, 6}$ and
$\itvllo{6,14}$ need to be updated. However, since $Zmin_{GM} = 6$ is
already larger than the recorded function values $3$ and $5$, we will
keep the original function values. A subsequent Intersects query will
start from either $Zmin = 3$ or $Zmin = 5$, which will be able to find
the inserted geometry $\itvl{6, 8}$. If an inserted geometry has
z-value larger than the maximum value in the piecewise function, we
will append a new piece at the end of the function.

For deletion, we do not perform any update because the piecewise
function would still provide lower bounds of the lowest intersecting
Z-addresses. However, we can end up with unnecessary refinements if
there are many deletions.  Hence, we periodically rebuild the
piecewise function using {\indextitle} if the lower bounds are too
loose and the number of refinements on average becomes too large.
Since deletion is less frequent than insertion in a typical workload,
it is not unreasonable to amortize the rebuild cost across a large
number of deletions.

	\section{Experiments}
\label{sec:experiments}
This section presents the result of an experimental analysis on {\indextitle} on query-only workloads, maintenance-only workloads and hybrid workloads (depicted in Section~\ref{subsec:hybrid} in the interest of space). All experiments are done in the main memory of a machine with 12th Gen Intel Core i9 CPU, 128GB memory, 1TB SSD storage.


\begin{table}
	\normalsize
	\caption{Dataset description}
	\vspace{-10pt}
	\resizebox{\linewidth}{!}{
		\begin{tabular}{|p{0.4\linewidth}|p{0.1\linewidth}|p{0.18\linewidth}|p{0.14\linewidth}|p{0.55\linewidth}|}
			\hline
			\textbf{Name }&
			\textbf{Size} &
			\textbf{Cardinality} &
			\textbf{Type} &
			\textbf{Description}   \\ \hline
			LINEARWATER (LW)~\cite{Tiger} &
			6.44GB &
			5.8M &
			LineString &
			Paths of rivers in the USA\\ \hline
			Roads~\cite{Tiger}                   & 8.29GB & 19M    &     LineString &  Paths of roads in the USA \\ \hline
			Parks~\cite{OSM} &
			8.53GB &
			9.8M &
			Polygon &
			Boundaries of parks and green areas on the planet \\ \hline
		\end{tabular}%
	}
	\label{tab:datasets}
	\vspace{-10pt}
\end{table}
\subsection{Experiment setup}


\textbf{Implementation details.}
We implement {\indextitle} on top of ALEX using C++ since ALEX is open-source and supports data updates.  Our implementation re-uses the hierarchical model and gapped arrays from ALEX and keeps the corresponding ALEX parameters unchanged.  
With that being said, {\indextitle} can be easily migrated to other learned indexes too.  Furthermore, our idea is also compatible with other one-dimensional index structures, such as B-trees.
We illustrate this by effortlessly implementing {\indextitle} on top of a B-tree, which incurs a minimal overhead with promising performance.

\textbf{Compared approaches.}
(1) Boost-Rtree: from Boost C++ v1.73.0, with default settings. (2) Quad-Tree:  from GEOS v3.9.0. Quad-Tree.(3) {\indextitle}-ALEX: This is the approach proposed in this paper. When querying for the $Contains$ relationship, no query augmentation is needed. However, for the $Intersection$ relationship, query augmentation comes into play.(4) {\indextitle}-BTREE: This uses the same approach as {\indextitle}-ALEX but with TLX-BTree as the base index, demonstrating the adaptability of {\indextitle} and the benefits derived from a one-dimensional index structure.
\begin{figure*}
	\begin{minipage}{.29\textwidth}
		\includegraphics[width=\linewidth]{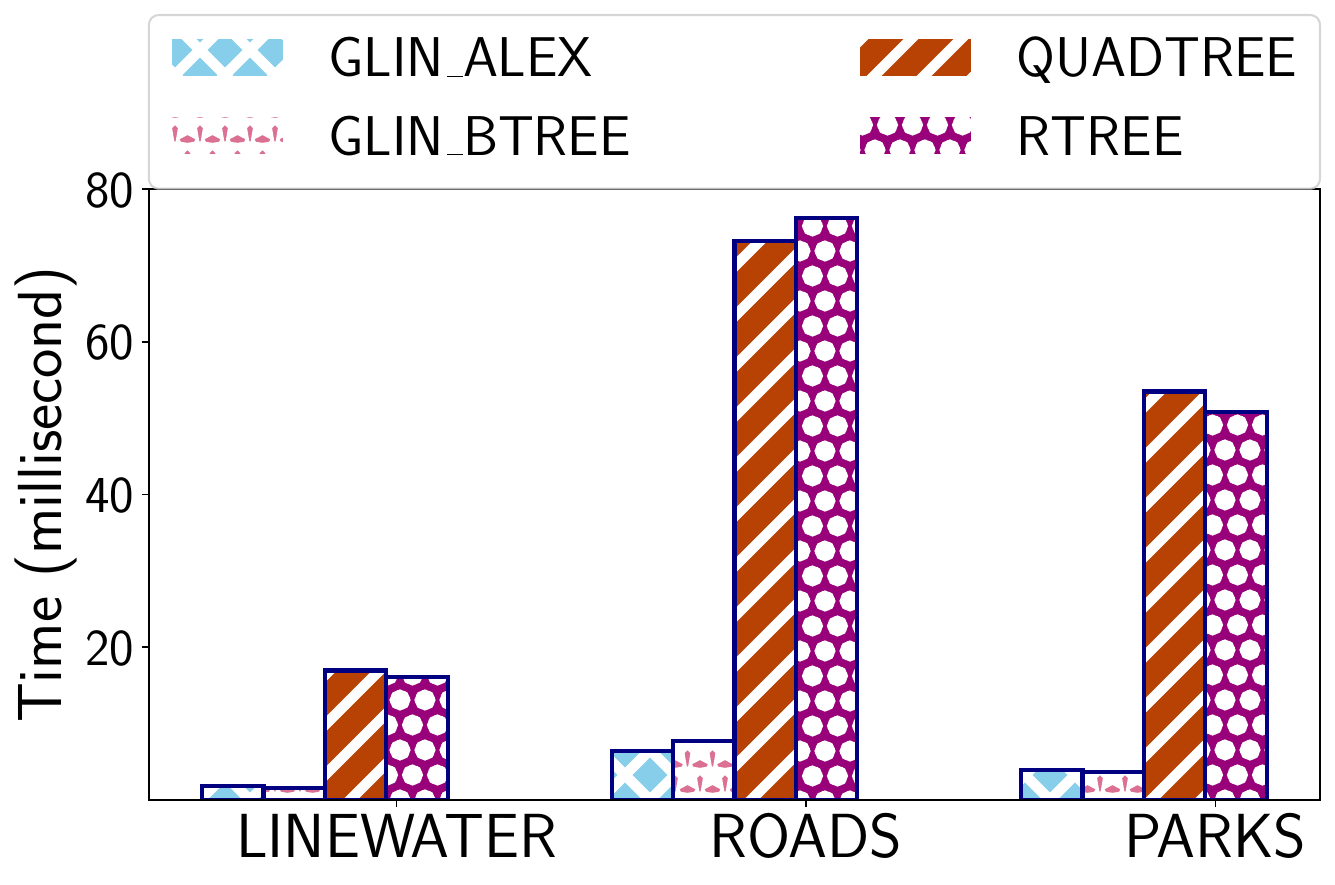}
		\subcaption{1\% Selectivity}
	\end{minipage}
	\begin{minipage}{.29\textwidth}
		\includegraphics[width=\linewidth]{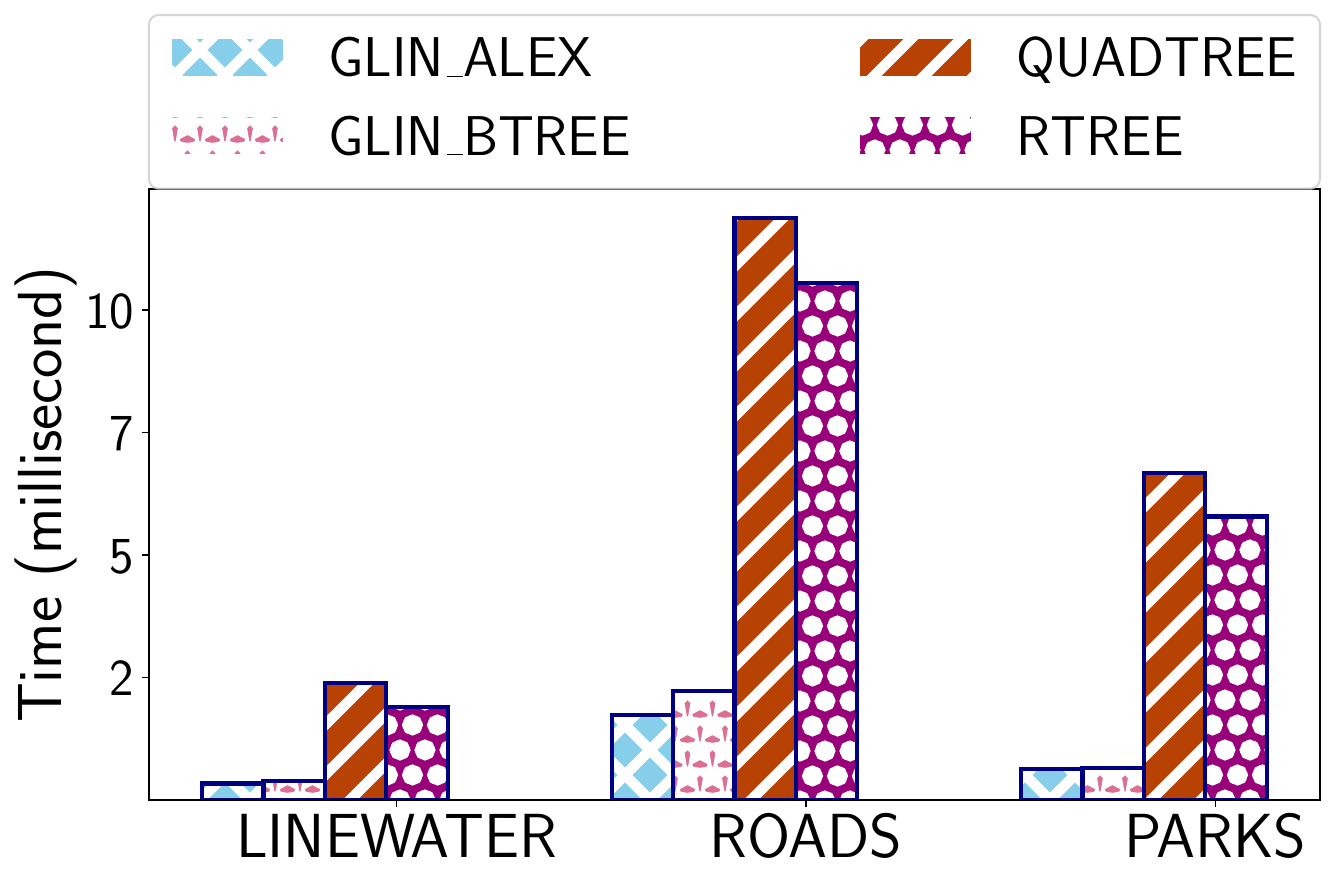}
		\subcaption{0.1\% Selectivity}
	\end{minipage}
	\begin{minipage}{.29\textwidth}
		\includegraphics[width=\linewidth]{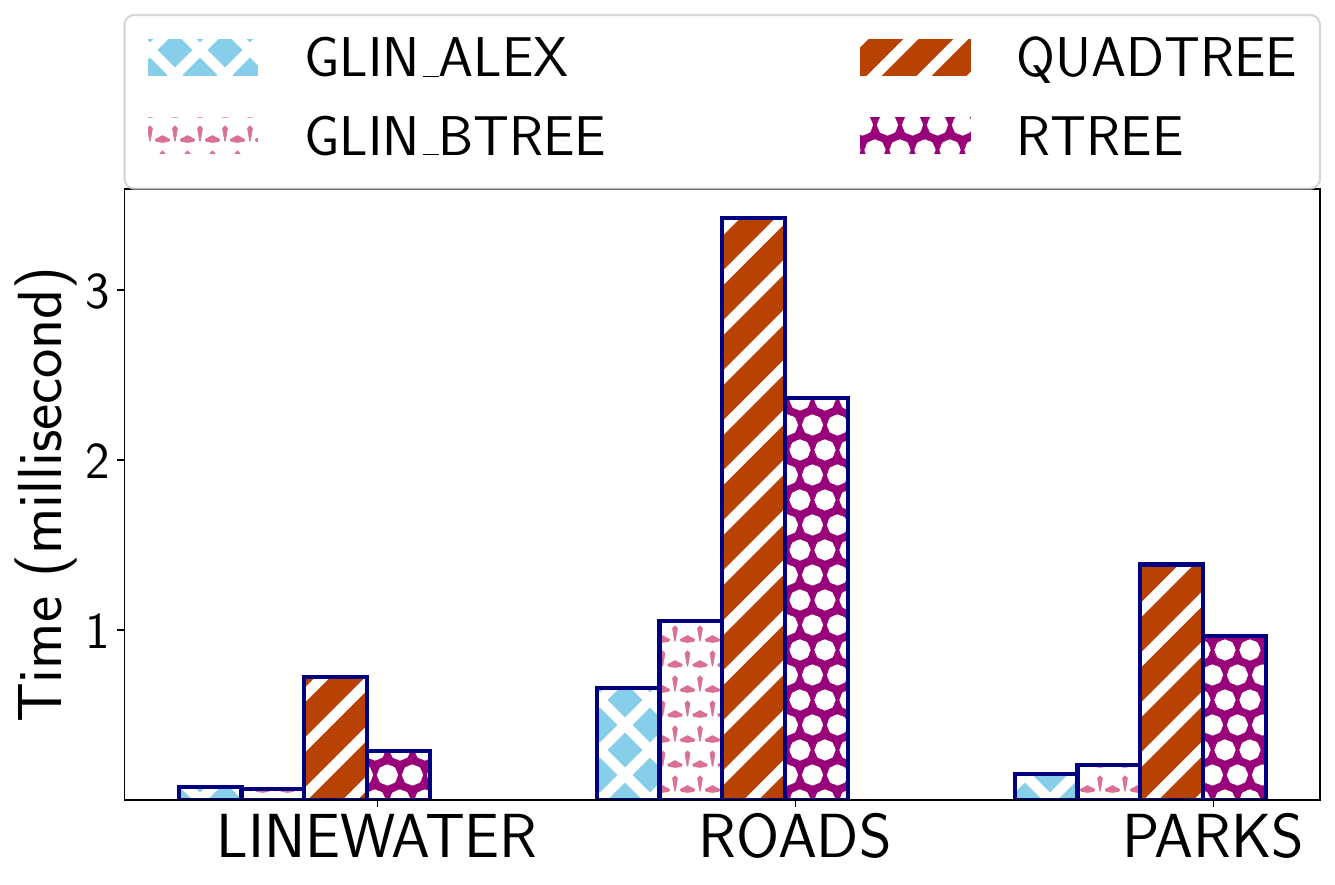}
		\subcaption{0.01\% Selectivity}
	\end{minipage}
	\caption{Query response time on different query selectivities with $Contains$ relationship}
	\label{fig:real-nopiece-overalltime}
	\vspace{-3mm}	
\end{figure*}
\begin{figure*}
	\begin{minipage}{.29\textwidth}
		\includegraphics[width=\linewidth]{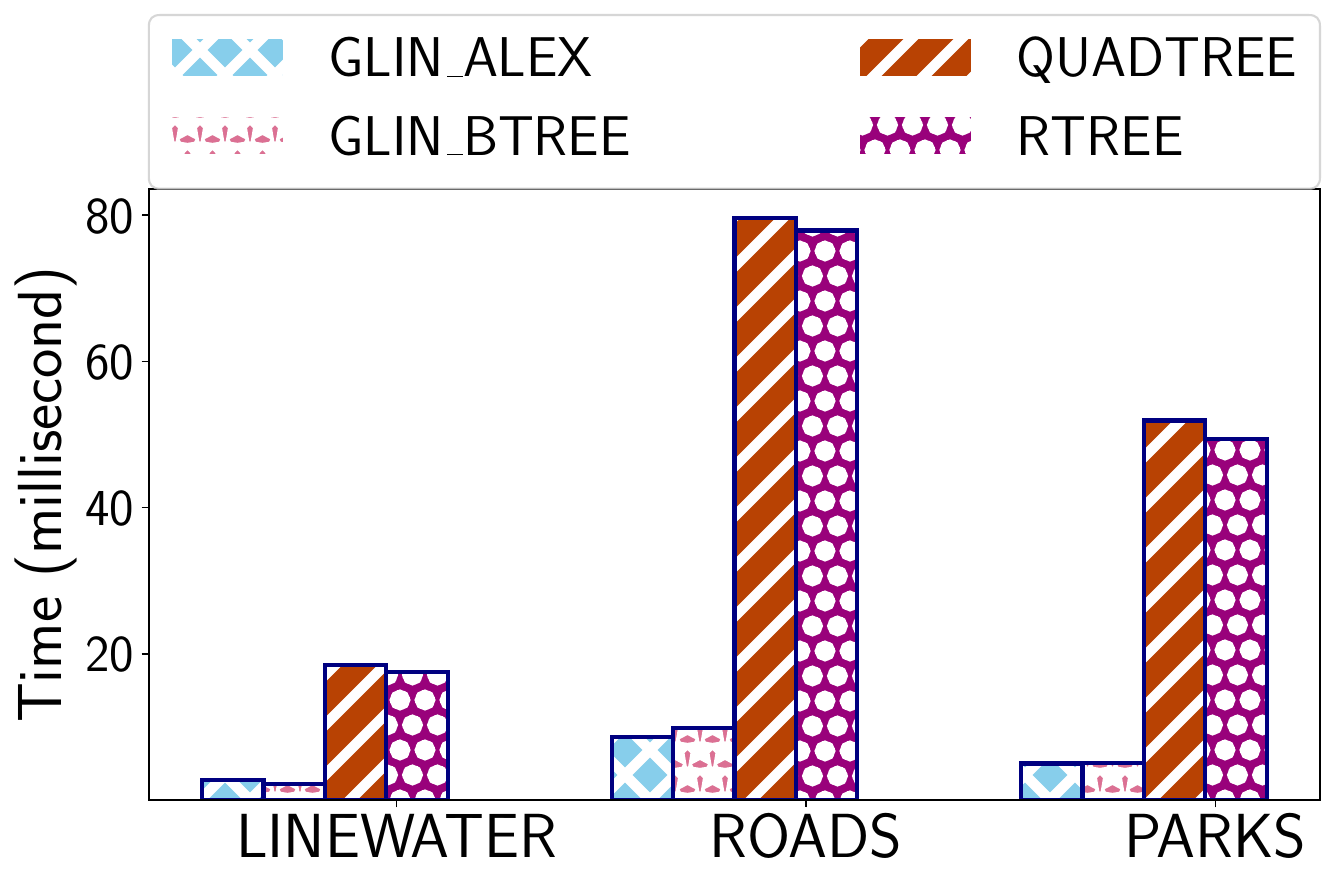}
		\subcaption{1\% Selectivity}
		\label{subfig:real-overalltime-01}
	\end{minipage}
	\begin{minipage}{.29\textwidth}
		\includegraphics[width=\linewidth]{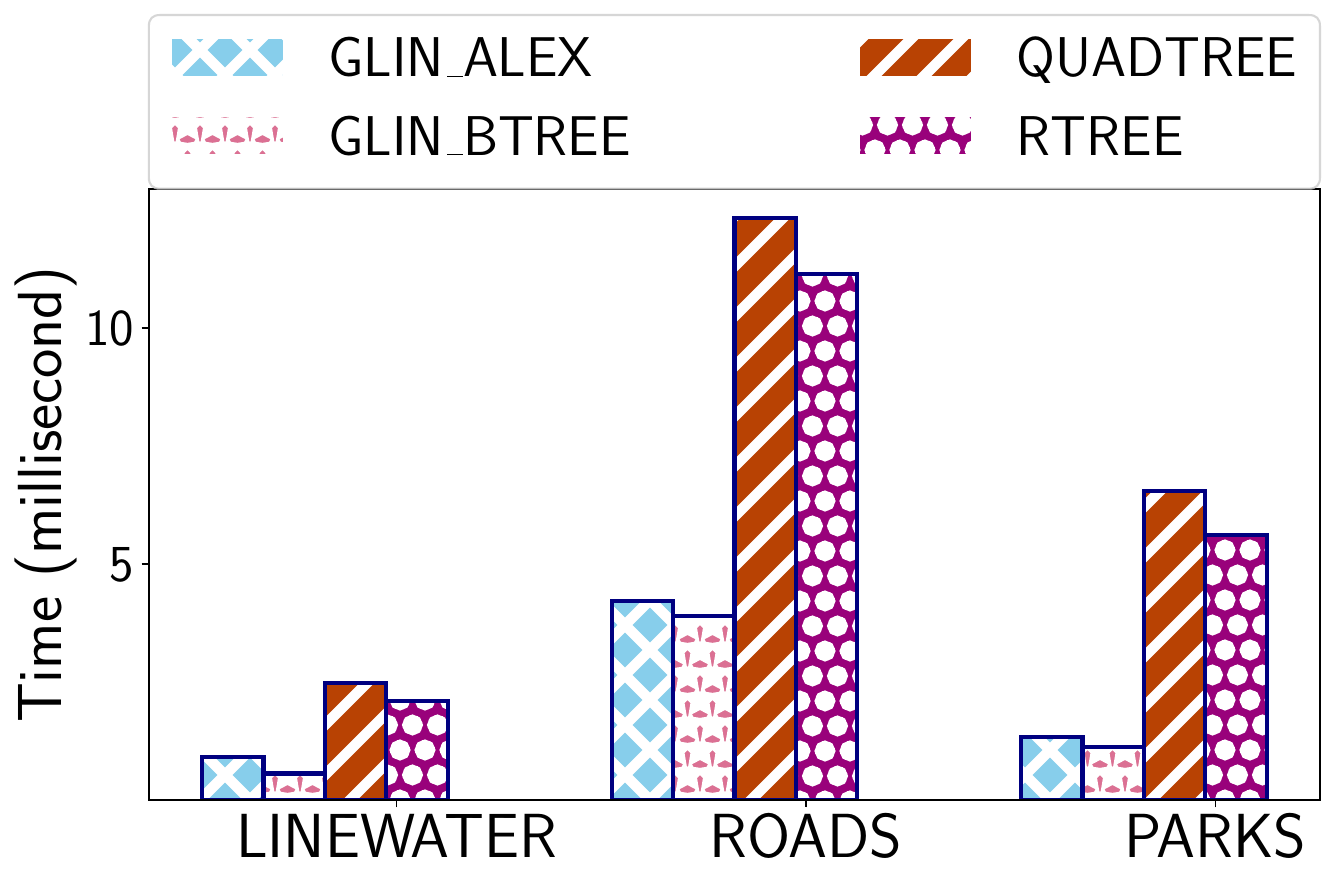}
		\subcaption{0.1\% Selectivity}
	\end{minipage}
	\begin{minipage}{.29\textwidth}
		\includegraphics[width=\linewidth]{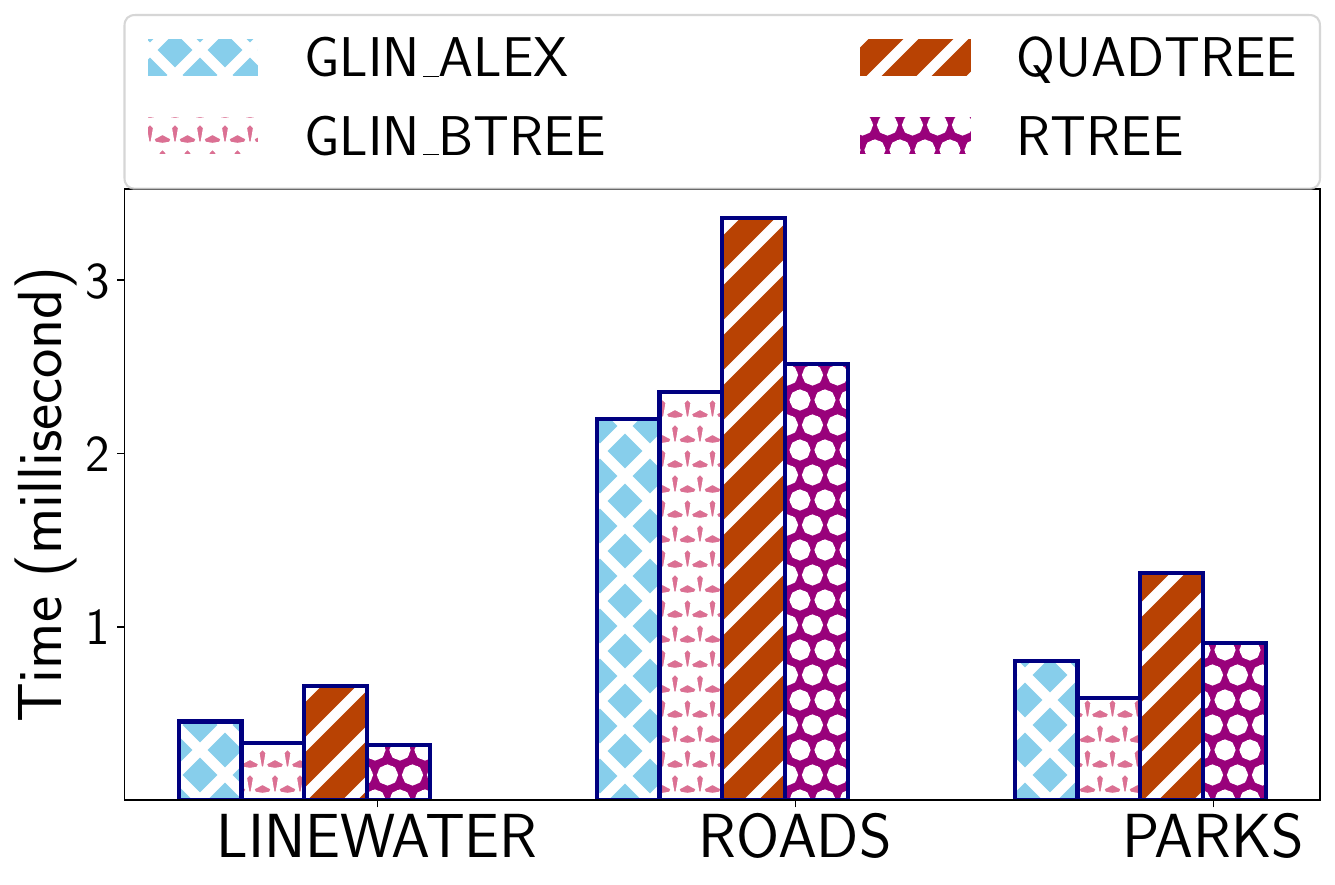}
		\subcaption{0.01\% Selectivity}
		\vspace{-3mm}	
	\end{minipage}
	\caption{Query response time on different query selectivities with $Intersects$ relationship}
	\label{fig:real-overalltime}
	\vspace{-3mm}	
\end{figure*}
\textbf{Datasets.}
We test our approaches on 3 real-world datasets(see Table~\ref{tab:datasets}), including polygon and line string data. Real-world datasets are obtained from the US Census Bureau TIGER project~\cite{Tiger} . These datasets are cleaned by SpatialHadoop~\cite{EM+15}.

\textbf{Query selectivity.}
We test {\indextitle} on 3 range query selectivities: 1\%, 0.1\%, 0.01\%. To generate a range query with the required selectivity, we randomly take a geometry from the dataset and do a K Nearest Neighbor query around this geometry (K = selectivity * dataset cardinality) using JTS STR-Tree The MBR of the KNN query results then becomes the query window at this selectivity. We generate 100 such query windows per selectivity per dataset.

\textbf{Query response time.}
The measured query response time consists of two parts: (1) index prob time. For all compared approaches, this is the time spent on searching the index structure. For {\indextitle}-ALEX and {\indextitle}-BTREE, this also includes the query augmentation time when check $Intersects$. (2) Refinement time. For all compared approaches, this is the time spent on refining the query results using the exact shapes of query windows and geometries. For the $Contains$ relationship, the results are refined using the $Contains$ check in GEOS while for the $Intersects$ relationship, the refinement uses the $Intersects$ check.

 \begin{figure}
	\includegraphics[width= 0.6\linewidth]{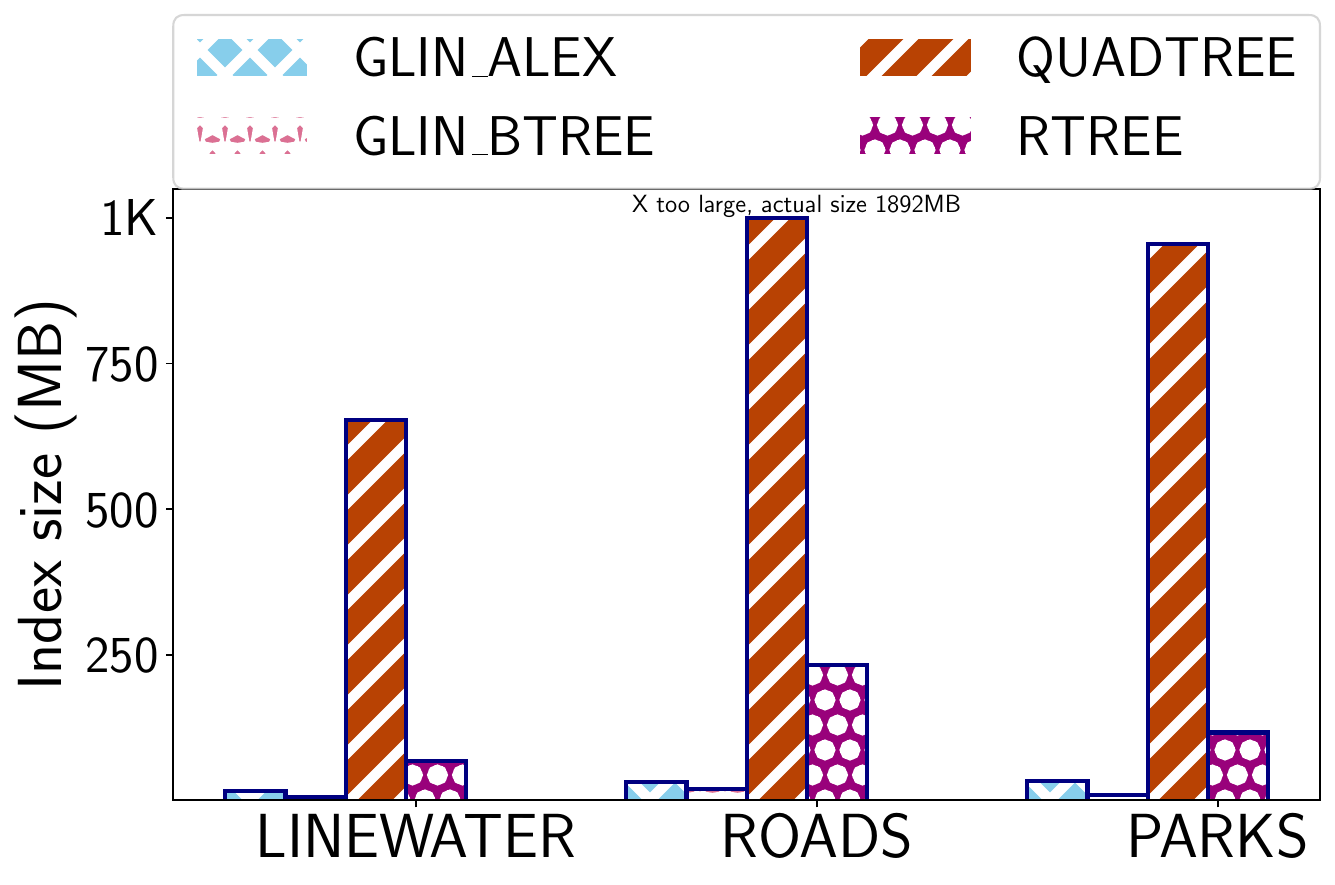}
		\vspace{0pt}
	\caption{Index size with the piecewise function}
	\label{fig:indexsize}
	\vspace{-10pt}
\end{figure}

\begin{figure}
	\includegraphics[width=0.7\linewidth]{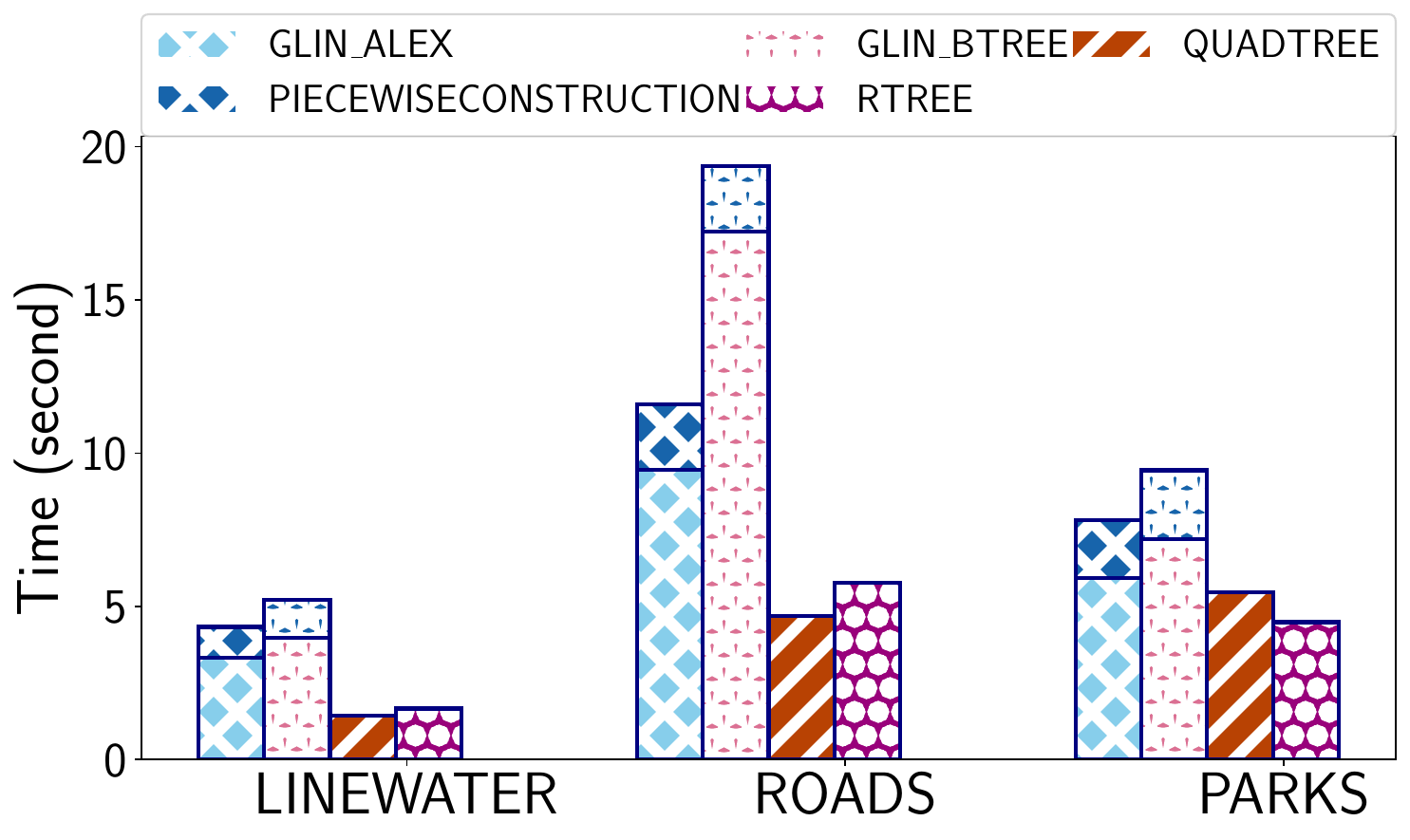}
	\caption{Index initialization time}
	\label{fig:initializationtime}
	\vspace{-10pt}
\end{figure}
\vspace{0pt}

\subsection{Query response time}

\textbf{Query response time for $Contains$.}
As shown in Figure~\ref{fig:real-nopiece-overalltime}, on 1\% - 0.01\% selectivity, the index probing time of {\indextitle}-ALEX and {\indextitle}-BTREE is nearly  30\% - 80\%shorter than Quad-Tree and R-Tree. On 0.01\% selectivity, {\indextitle}-ALEX and {\indextitle}-BTREE are still 1 times to 3 times faster than R-Tree and Quad-Tree. This makes sense because {\indextitle}-ALEX uses the model prediction-based traversal and {\indextitle}-BTREE uses pointer to traverse while Quad-Tree and R-Tree do the comparison-based tree traversal. When check $Contains$, there is no need to perform the query augmentation.

\textbf{Query response time for $Intersects$.}
Figure~\ref{fig:real-overalltime} demonstrates the query performance of both {\indextitle}-ALEX and {\indextitle}-BTREE when they incorporate query augmentation to perform query with the $Intersects$ relationship. Performing query augmentation in {\indextitle} introduces additional overheads during index construction and querying. These overheads include: an additional traversal of the leaf level to construct the piecewise function, the piecewise function to augment the query window, a small auxiliary index (either ALEX or BTREE) to handle outliers, and an extra search on the piecewise function when augmenting the query window. Despite these overheads, as evidenced by Figure~\ref{fig:real-overalltime}, they do not significantly impact the query performance. Both {\indextitle}-ALEX and {\indextitle}-BTREE can achieve a query performance comparable to that of {\indextitle} when it does not utilize query augmentation.

\begin{figure*}
	\begin{minipage}{.29\textwidth}
		\includegraphics[width=\linewidth]{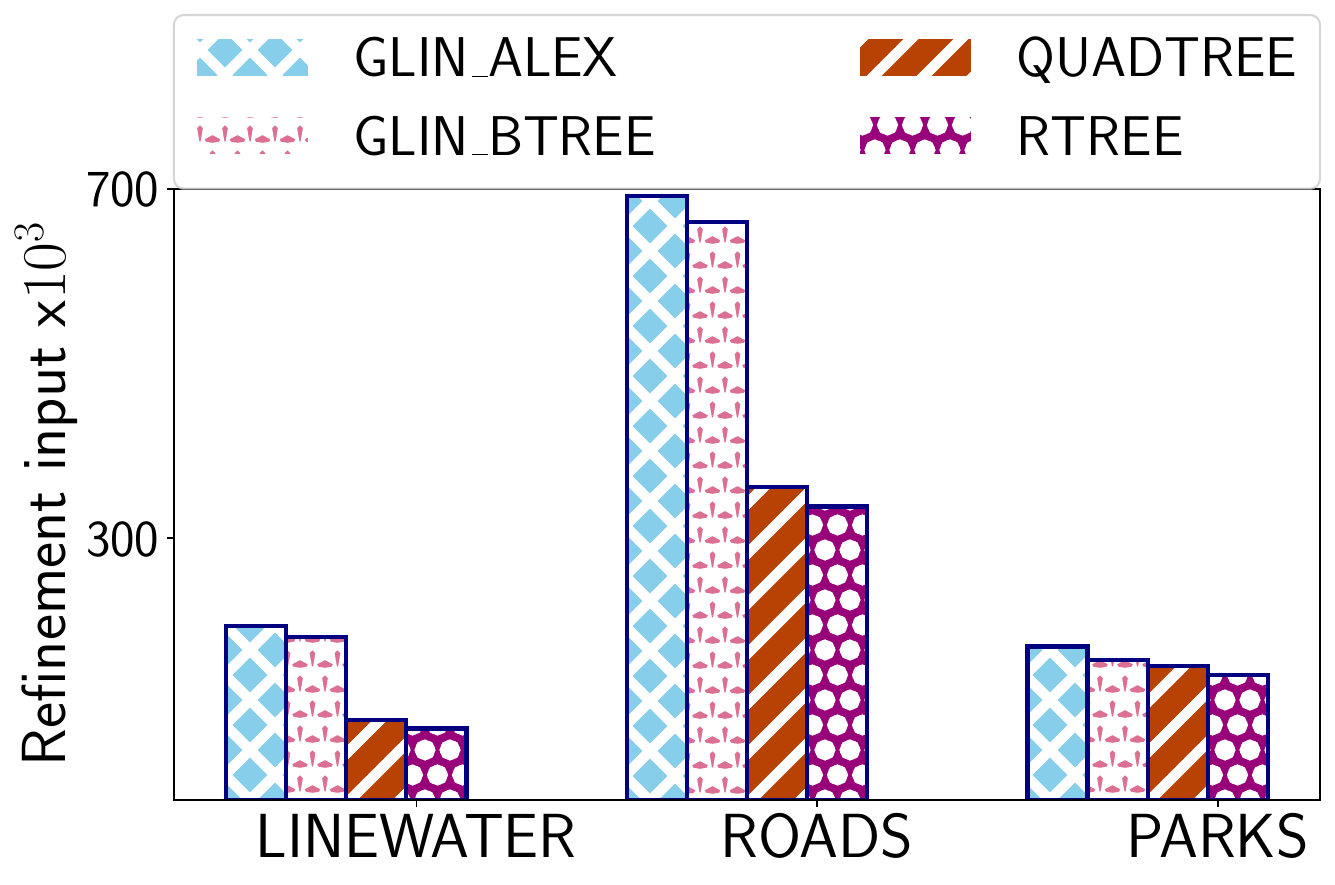}
		\subcaption{1\% selectivity}
	\end{minipage}
	\begin{minipage}{.29\textwidth}
		\includegraphics[width=\linewidth]{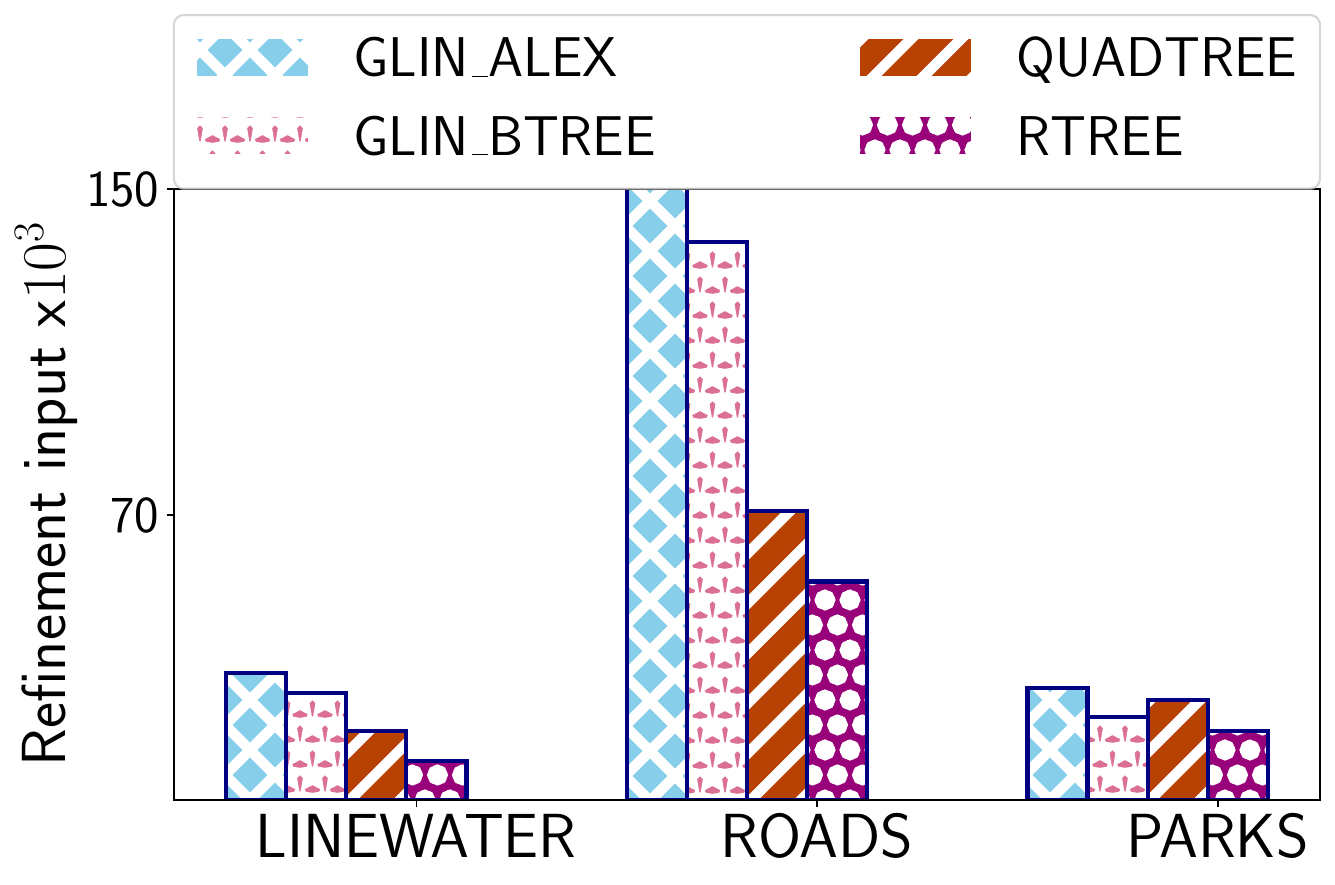}
		\subcaption{0.1\% selectivity}
	\end{minipage}
	\begin{minipage}{.29\textwidth}
		\includegraphics[width=\linewidth]{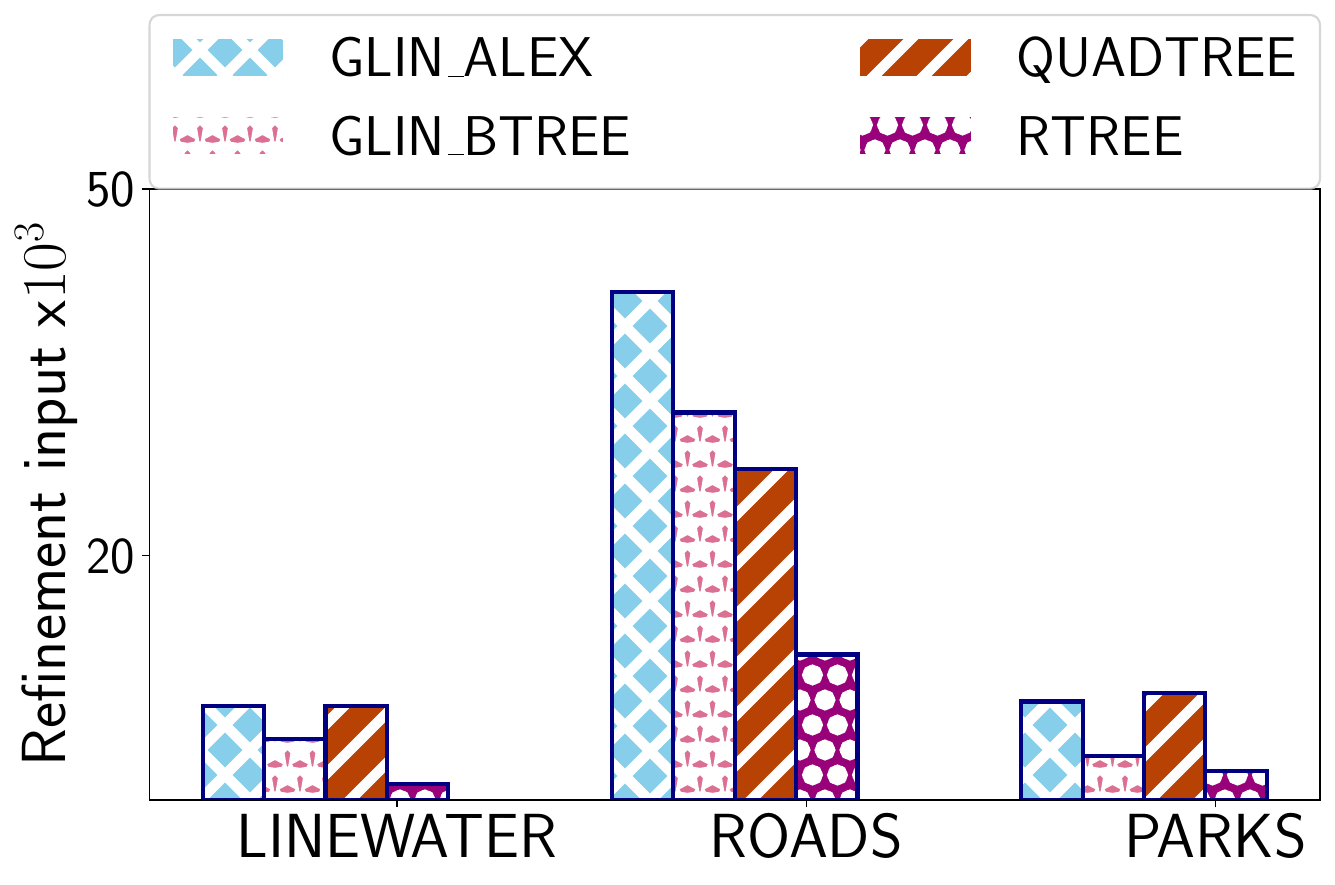}
		\subcaption{0.01\% selectivity}
	\end{minipage}
	\caption{Number of records checked during the refinement}
	\label{fig:real_num_before_refine}
	\vspace{-3mm}	
\end{figure*}

\textbf{False positives.}
Figure~\ref{fig:real_num_before_refine} illustrates the number of records checked during the refinement. A lower value indicates less false positives which eventually leads to less refinement time and overall query response time. {\indextitle} has 10\% - 50\% more false positives than R-Tree on 1\% to 0.1\% selectivity. But the fast index probing of {\indextitle} makes up the difference so its query response time still outstanding to others. As expected, {\indextitle} with query augmentation introduces more false positives. the larger the selectivity, the more number of record to check during the refinement.
\subsection{Indexing overhead}

{\bf Index size.}
We measure the sizes by combining the size of internal nodes and the size of leaf node metadata. As demonstrated in Figure ~\ref{fig:indexsize}, the index size of {\indextitle} is 80\% - 90\% smaller than that of the Quad-Tree, and 60\% - 80\%times smaller than the R-Tree when tested on real-world datasets. This is reasonable considering that {\indextitle} has far fewer nodes, and each internal node employs a simple linear regression model comprised only of two parameters. Additionally, we calculated the sizes of {\indextitle}-ALEX and {\indextitle}-BTREE, including the piecewise function, and found that this part is very small, so it does not alter our conclusion.

{\bf Index initialization time.}
As depicted in Figure ~\ref{fig:initializationtime}, {\indextitle} requires 10\% - 50\% more initialization time compared to Quad-Tree and R-Tree on real-world datasets. This is understandable as, during index initialization, {\indextitle} needs to sort geometries by their $Zmin$ values and train models. {\indextitle} with query augmentation takes approximately 10\% more time than {\indextitle} because it needs to generate the piecewise function and handle outliers. The top part of {\indextitle}-ALEX and {\indextitle}-BTREE, depicted in a deeper color, represents the construction time of the piecewise function and additional auxiliary index for query augmentation, which is not significantly more than the original construction time of {\indextitle}.
\subsection{Tuning {\indextitle} parameters}

This section studies the impact of the piece\_ granularity (m) parameter. This parameter defines the number of records summarized by each piece of the piecewise function. 
\begin{figure}
	\includegraphics[width=0.7\linewidth]{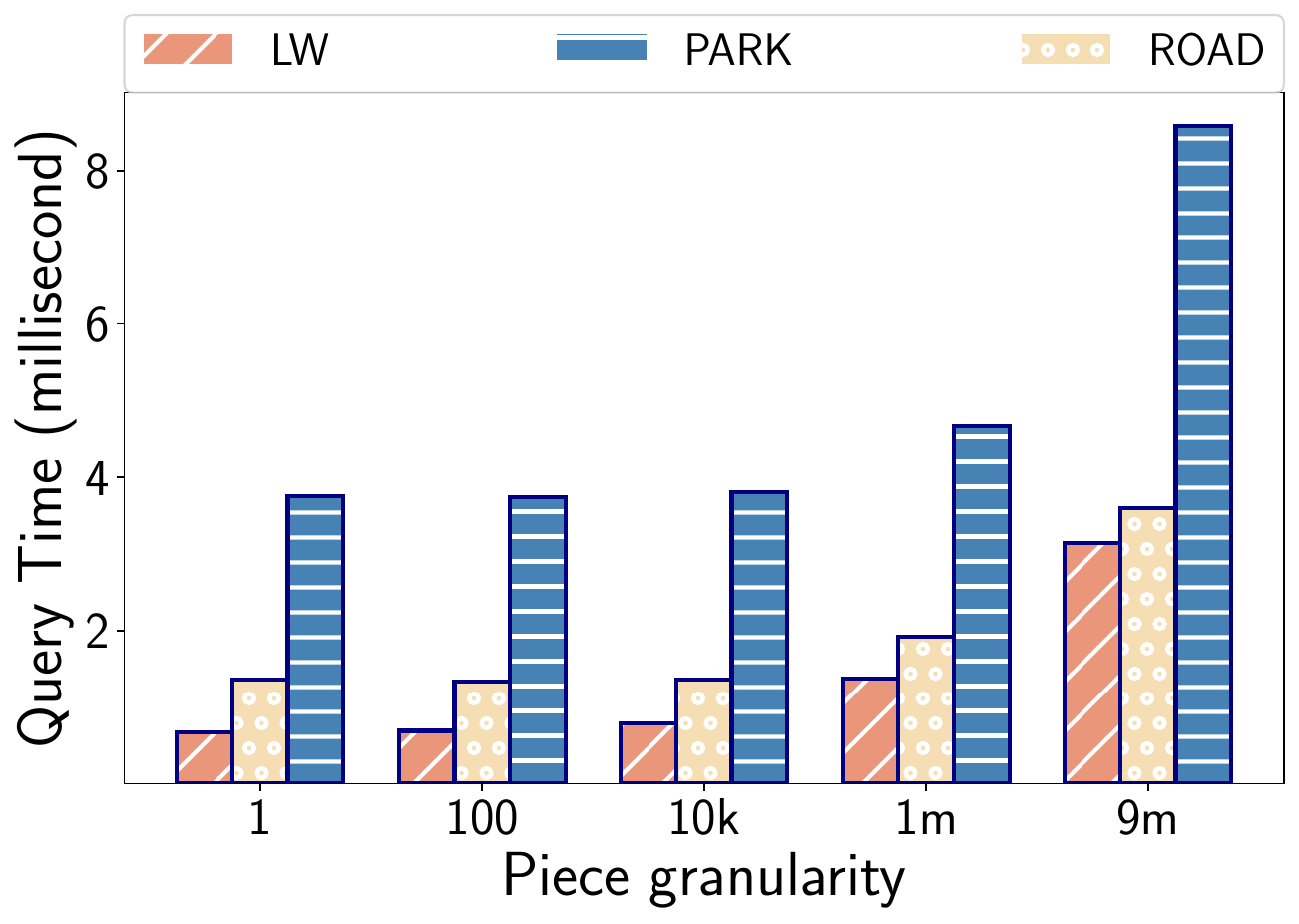}
	\vspace{0pt}
	\caption{{\indextitle} query time on piece granularity ($Intersects$)}
	\label{fig:para-overall}
	\vspace{-3mm}
\end{figure}

\textbf{Index probing time.}
As shown in Figure~\ref{fig:para-overall}, piece\_granularity has a significant impact on the index probing time.  The probing time for piece\_granularity = 9 million is up to twice as high as that for piece\_granularity = 10000.  A larger piece\_granularity results in higher index probing time because it can augment every query window to a small $Zmin$. This may potentially cause every query to start from the beginning of the leaf level, resulting in a time-consuming leaf level scan and more false positive refinement. The condition for piece\_granularity = 9 million illustrates this scenario. Conversely, a smaller piece\_granularity doesn't significantly reduce the query response time. Even though the query window might not be augmented to a small $Zmin$, leading to more refinement, the search within the piecewise function will take longer for each augmented query window. As a result, the query response time for piece\_granularity = 1 is not significantly smaller than the time for larger granularities, such as 10000.

\textbf{Index size.}
Compared to the index size of {\indextitle} (see Figure~\ref{fig:indexsize}), the storage overhead of the piecewise function is negligible. Each piece contains a $Zmax$, indicating where the current piece ends, and a $Zmin$, used to augment a query window if it falls within this piece. As such, the storage overhead of the piecewise function is insubstantial. Moreover, Figure~\ref{fig:indexsize} also incorporates the auxiliary index, yet the overall size remains significantly smaller than those of QUADTREE and RTREE. Therefore, we can conclude that the storage overhead of {\indextitle} is indeed negligible.

Therefore, we use piece\_granularity = 10000 as the default parameter as it shows the good performance in Figure~\ref{fig:para-overall} and its piecewise function size is very small compared to {\indextitle} index size.


\subsection{Maintenance Overhead}

\begin{figure}
	\begin{minipage}{.3\textwidth}
		\includegraphics[width=\linewidth]{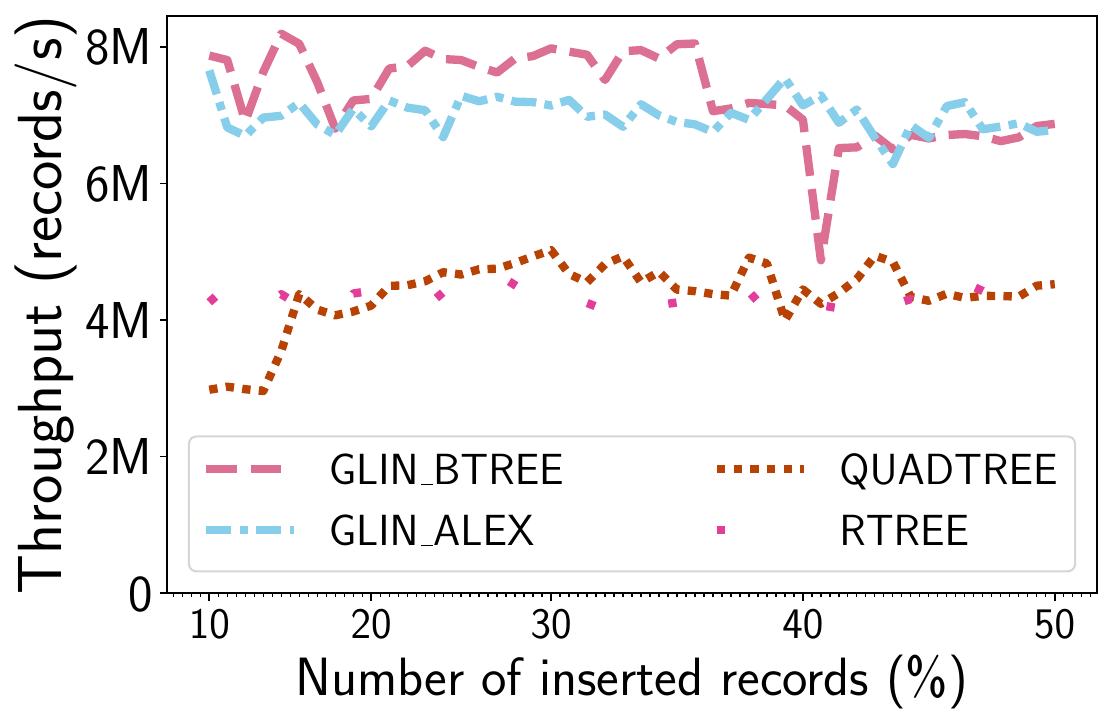}
		\label{subfig:insertion}
	\end{minipage}
	\begin{minipage}{.3\textwidth}
		\includegraphics[width=\linewidth]{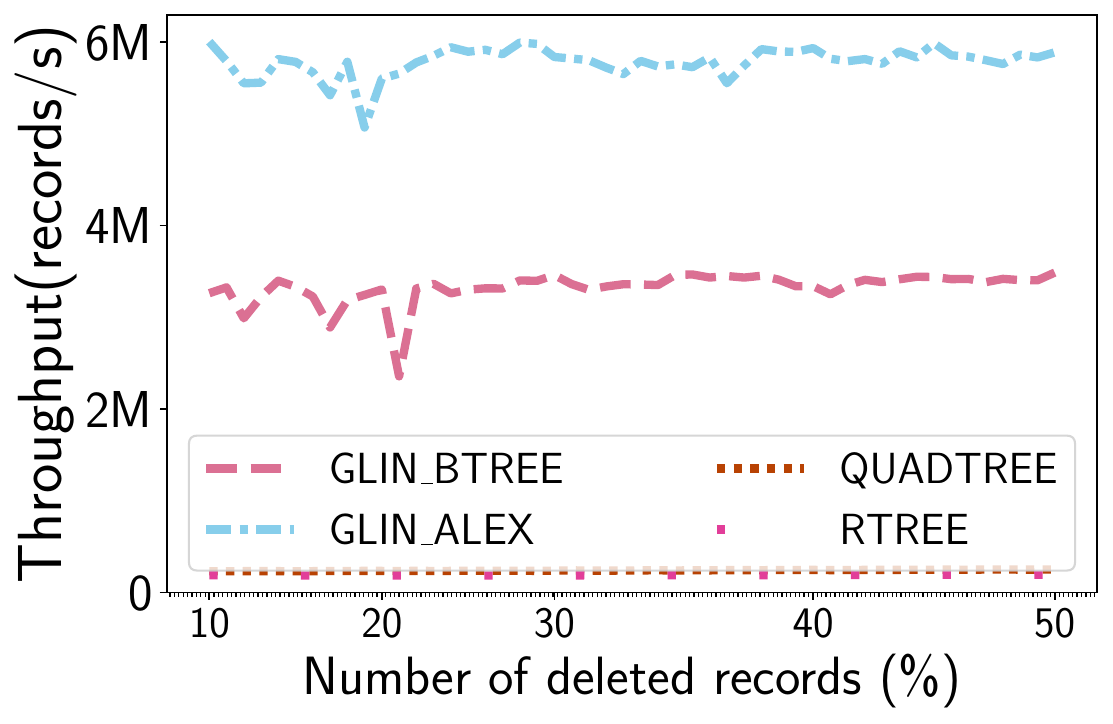}
		\label{subfig:deletion}
	\end{minipage}
	\caption{Index maintenance performance on ROADS}
	\label{fig:realinsertion}
	\vspace{-10pt}	
\end{figure}

{\bf Insertion.}
For each dataset, we first bulk-load a random 50\% of the data into all indexes and then insert the remaining 50\% into the indexes record by record. As shown in Figure~\ref{fig:realinsertion}(a), on larger datasets, the throughput of {\indextitle} is around 1.5 times higher than R-Tree and 1.2 times higher than Quad-Tree. This makes sense because {\indextitle}'s index probing is orders of magnitude faster than others  and no refinement is needed for insertion. {\indextitle} occasionally show performance downgrade because of node expansion or splitting. 

{\bf Deletion.}
For each dataset, we first bulk-load the entire dataset and then randomly delete 50\% of the data record by record. As shown in Figure~\ref{fig:realinsertion}(b),  the throughput of {\indextitle} is around 3 - 5 times higher than R-Tree and Quad-Tree as {\indextitle}'s index probing is orders of magnitude faster than others. The throughput of{\indextitle} occasionally show performance downgrade because of node merging.

	\section{Related Work}
\label{sec:related-work}

\textbf{Learned indexes.}
Recursive model index(RMI)~\cite{KBC+18}, a tree-like hierarchical model, to learn the cumulative distribution function(CDF) between keys and their position. RMI takes as input a lookup key and predicts the corresponding position by using the models level by level. Compared to B+ Tree, RMI possesses low storage overhead with an outperforming lookup performance but only supports read-only workload. Hermit~\cite{WYT+19} is a learned secondary index that leverages a hierarchical machine learning model to learn the correlation between two columns. ALEX~\cite{DMY+20} is an updatable learned index which adopts RMI's hierarchy structure but adds gapped arrays and node splitting to absorb data updates.

\textbf{Learned spatial indexes.}
Researchers have been working on extending learned indexes to uphold spatial and multi-dimentional point data. 
ZM-index\cite{WFX+19} leverages the Z-order space-filling curve to sort the data and then builds RMI on them. Given a spatial range query, it first maps a range query to two Z addresses, then uses the prebuilt machine learning model to find an approximate range for further investigation. Although both {\indextitle} and ZM-Index make use of Z order curve, ZM-index cannot handle non-point data and only works with read-only workload.  Flood~\cite{NDA+20} also employs the RMI to support multidimensional data. It proposes an in-memory read optimized index that partitions a d-dimensional space with a d-1 dimensional grid. The model will predict the grid cell that contains the lookup key. The ML-Index\cite{DMM+20} utilizes the iDistance~\cite{JOT+05} to map data points to the one-dimensional value and also employs the RMI to index the values further. Qi et al.~\cite{QLJ+20} come up with a recursive spatial model index called RSMI to improve the ZM-index. Their work mitigates the uneven gap problem by using a rank space-based transformation. However, this work provides approximate answers. Li et al. propose LISA~\cite{LLZ+20}, a disk-based learned spatial index that can reach a low storage consumption and I/O cost. LISA partitions the space to grids and assigned each grid an ID by applying the partially monotonic function. All of the works mentioned above focus on making learned indexes work for 2 or multi-dimensional point data. Unfortunately, in the real world, geospatial data is more than just points. {\indextitle} handles all types of geometries and hence is a practical alternative to R-Tree or Quad-Tree.


\textbf{Lightweight index structures.}
Some other studies focus on succinct index structures which take advantage of data synopses from the indexed data and quickly skip irrelevant data. Column imprints~\cite{SK+13} utilizes the idea of cache conscious bitmap indexing to create a bit map for each zone. Block Range Indexes (BRIN) in Postgres stores min/max values for each range of disk blocks. Hippo~\cite{YS16} extends BRIN's idea but implements partial histograms in each range to decrease the query response time. Hentschel et al. propose Column Sketch\cite{HKI+18} that makes use of lossy compression to generate data synopses and hence accelerates table scan. BF-tree\cite{AA14} applies bloom filter in the leaf node of a B-Tree and hence reduces the storage overhead. However, these succinct index structures reduce the index storage overhead at the cost of additional query response time and cannot be easily tailored to complex geometries.
	\section{Conclusion}
\label{sec:conclusion}
This paper introduces {\indextitle}, a lightweight learned index for spatial range queries on complex geometries. In terms of storage overhead, {\indextitle} is 80\% - 90\% less than Quad-Tree and 60\% - 80\% times less than R-Tree. Moreover, {\indextitle}'s maintenance speed is around 1.5 times higher on insertion and 3 -5 times higher on deletion as opposed to R-Tree and Quad-Tree. If the application only needs the $Contains$ relationship, the user can opt to use {\indextitle} without query augmentation the query response time is 30\% - 80\% shorter than Quad-Tree and R-Tree on medium selectivity. {\indextitle} with query augmentation deals with both spatial relationships still showing a 30\%-70\% faster than Quad-Tree and R-Tree query response time on medium selectivity. In a nutshell, {\indextitle} is a lightweight indexing mechanism for medium selectivity queries which are commonly used in spatial analytic applications.


	\clearpage
	
	\appendix
	\section{APPENDIX}
\label{sec:proofs}

\subsection{Proofs of Lemma~\ref{lem:contains} and
Lemma~\ref{lem:intersects}}

To prove the lemmas, we first show a well-known result in
Theorem~\ref{theo:monotonic} and additionally prove
Theorem~\ref{theo:mbr}.

\begin{theo}\label{theo:monotonic}
	{\bf Monotonic ordering}~\cite{LZL+07}: Data points ordered by non-descending Z-addresses are monotonic in a way that a dominating point is placed before its dominated points. 
\end{theo}

where \emph{dominance} is defined as: given two points $p$ and $p'$, if $p$ is no larger than $p'$ in any dimension, then we say $p$ dominates $p'$.

\begin{theo}\label{theo:mbr}
	If Q contains GM, then MBR$_Q$ contains MBR$_{GM}$
\end{theo}
where MBR$_Q$ contains MBR$_{GM}$ $\iff$ Q's $p_{min}$ dominates GM's $p_{min}$ and GM's $p_{max}$ dominates Q's $p_{max}$.

\begin{proof}
	Since Q $contains$ GM, in a 2D space, if there is a line, it is obvious that the geometrical projection of Q on this line must contain the geometrical projection of GM on this line. In other words, when Q $contains$ GM, then if a person stands somewhere outside Q, he or she should never see GM because GM is completely inside Q assuming Q is a closed geometry. The projection of a geometry on X axis and Y axis are $[x_{min}, x_{max}]$ and $[y_{min}, y_{max}]$, respectively. We have (1) Q's $x_{min} \leq$ GM's $x_{min}$ \& Q's $y_{min} \leq$ GM's $y_{min}$, so Q's $p_{min}$ dominates GM's $p_{min}$ (2) GM's $x_{max} \leq$ Q's $x_{max}$ \& GM's $y_{max} \leq$ Q's $y_{max}$, so GM's $p_{max}$ dominates Q's $p_{max}$.
\end{proof}

Proof of Lemma~\ref{lem:contains}:
\begin{proof}
	$Zmin_{GM} \leq Zmax_{GM}$ is known. Since Q $contains$ GM, we have (1) $p_{min}$ of Q dominates $p_{min}$ of GM, so $Zmin_Q \leq Zmin_{GM}$ (2) $p_{max}$ of GM dominates $p_{max}$ of Q, so $Zmax_{GM} \leq Zmax_Q$.
\end{proof}

Proof of Lemma~\ref{lem:intersects}:
\begin{proof}
	Since Q $Intersects$ GM, Q and GM must share some portion of the space. On the other hand, Q and GM's $Zitvl$ guarantee to cover any point that falls inside Q and GM, respectively. Therefore, $Zitvl_Q$ and $Zitvl_{GM}$ share some portion of the intervals.
\end{proof}

\subsection{Monotonicity of piecewise functions}

\begin{lemma}
	The piecewise function for query augmentation is a non-strict
	monotonically increasing function.
\end{lemma}
\begin{proof}
To show that by contradition, suppose there are two $Zmax$ values
$Zmax_1 < Zmax_2$ but the lowest intersecting Z-addresses $f(Zmax_1) >
f(Zmax_2)$. Let the pieces containing $Zmax_1$ and $Zmax_2$ be
the $i_1$ and $i_2$. Since the computed $f$
values are different and $Zmax_1 < Zmax_2$, we must have $i_1 < i_2$.
For the second piece, there must be a z-address interval
$\itvl{f(Zmax_2), Zmax_2'}$ such that $Zmax_2' \in \itvl{Z_{i_2},
Z_{i_2+1}}$. Then we can show that the z-address interval
$\itvl{f(Zmax_2), Zmax_2'}$ intersects with the first piece:
\begin{eqnarray*}
	f(Zmax_2) & < & f(Zmax_1)  \qquad (\textrm{assumption}) \\
		&\leq& Z_{i_1+1} \qquad (\textrm{definition of }f)  \\
		&\leq& Z_{i_2} \qquad (\textrm{since } i_1 < i_2) \\
		&\leq& Zmax_2' \qquad (\textrm{definition of }f)
\end{eqnarray*}
	Therefore, $f(Zmax_1)$ should have been smaller than or
equal to $f(Zmax_2)$, which is a contradition.
\end{proof}

\subsection{Z-address cell size}
\label{subsec:cellsize}

\begin{figure}
	\centering
	\includegraphics[width=0.5\linewidth]{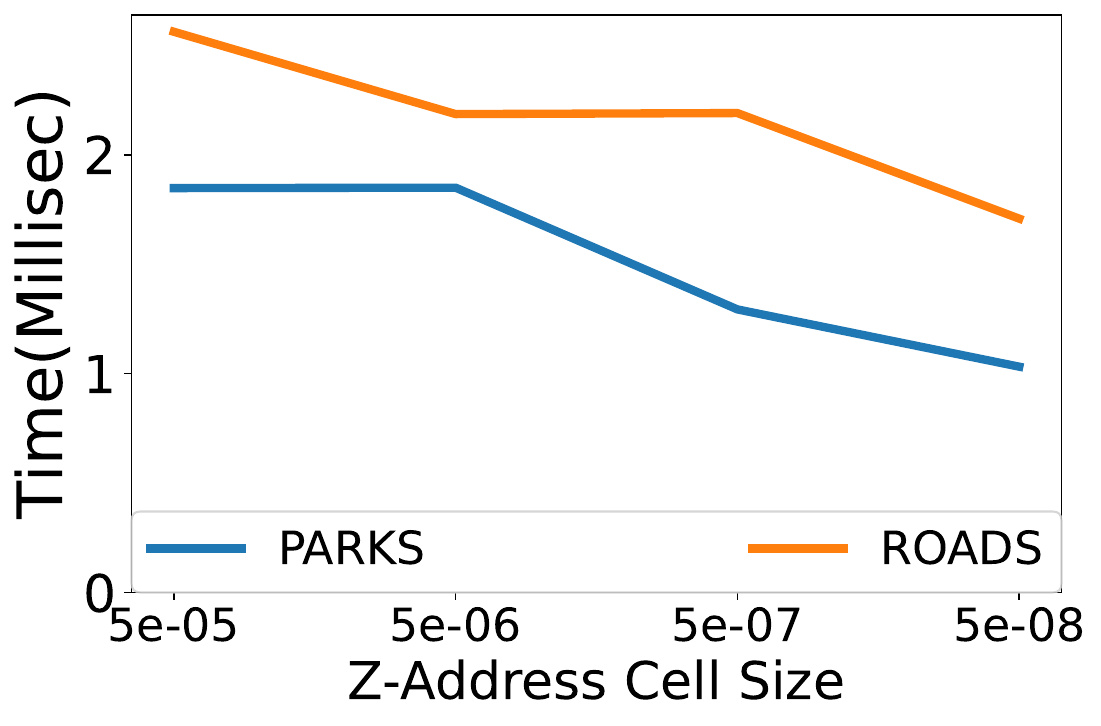}
	\caption{Query response time per cell size}
	\label{fig:cell-size}
\end{figure}

The cell size can be any small number with 7 to 8 decimal places (see
Figure~\ref{fig:cell-size}). This way, we can prevent
too many geospatial coordinates from having the same Z-address, which
would otherwise make indexe unable to prune the refinement candidates.

\subsection{{\indextitle}'s performance on hybrid workload}
\label{subsec:hybrid}

\begin{figure}
	\begin{minipage}{.3\textwidth}
		\includegraphics[width=\linewidth]{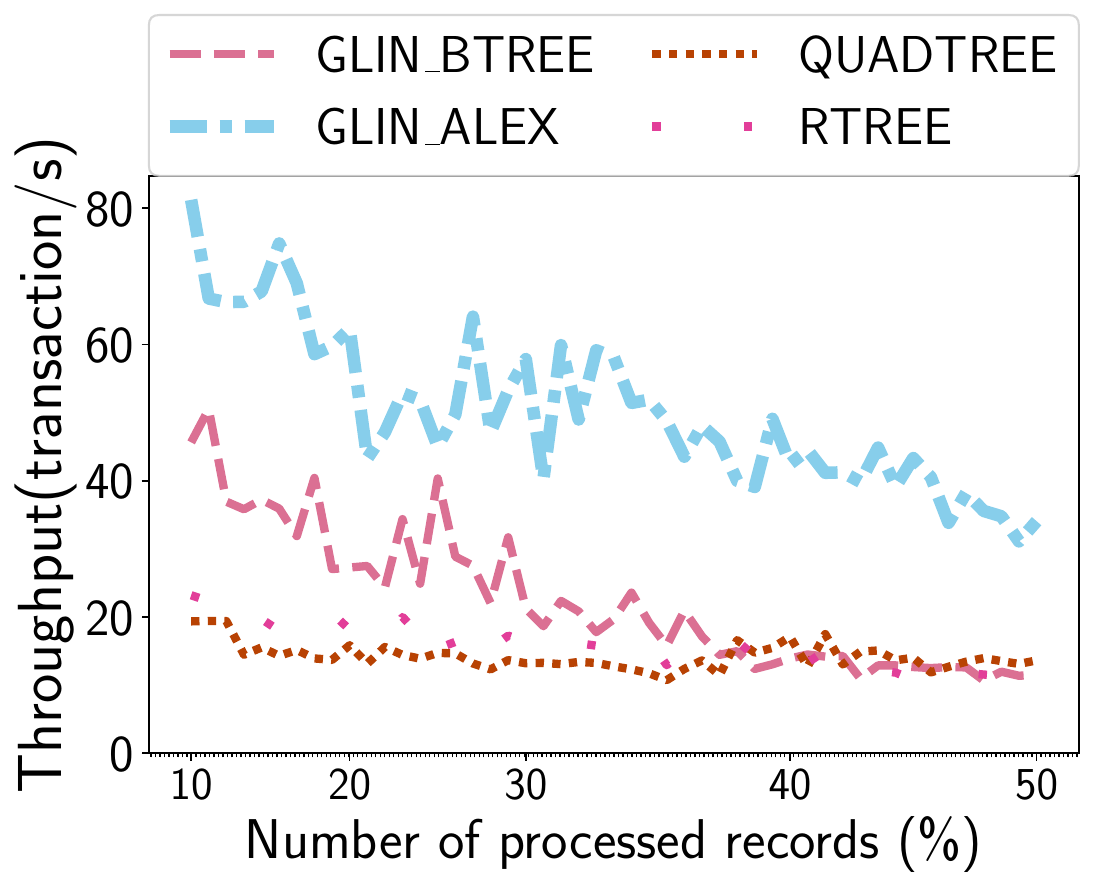}
		\subcaption{ Read-intensive on Roads}
		\label{subfig:read-intensive-roads}
	\end{minipage}
	
	\begin{minipage}{.3\textwidth}
		\includegraphics[width=\linewidth]{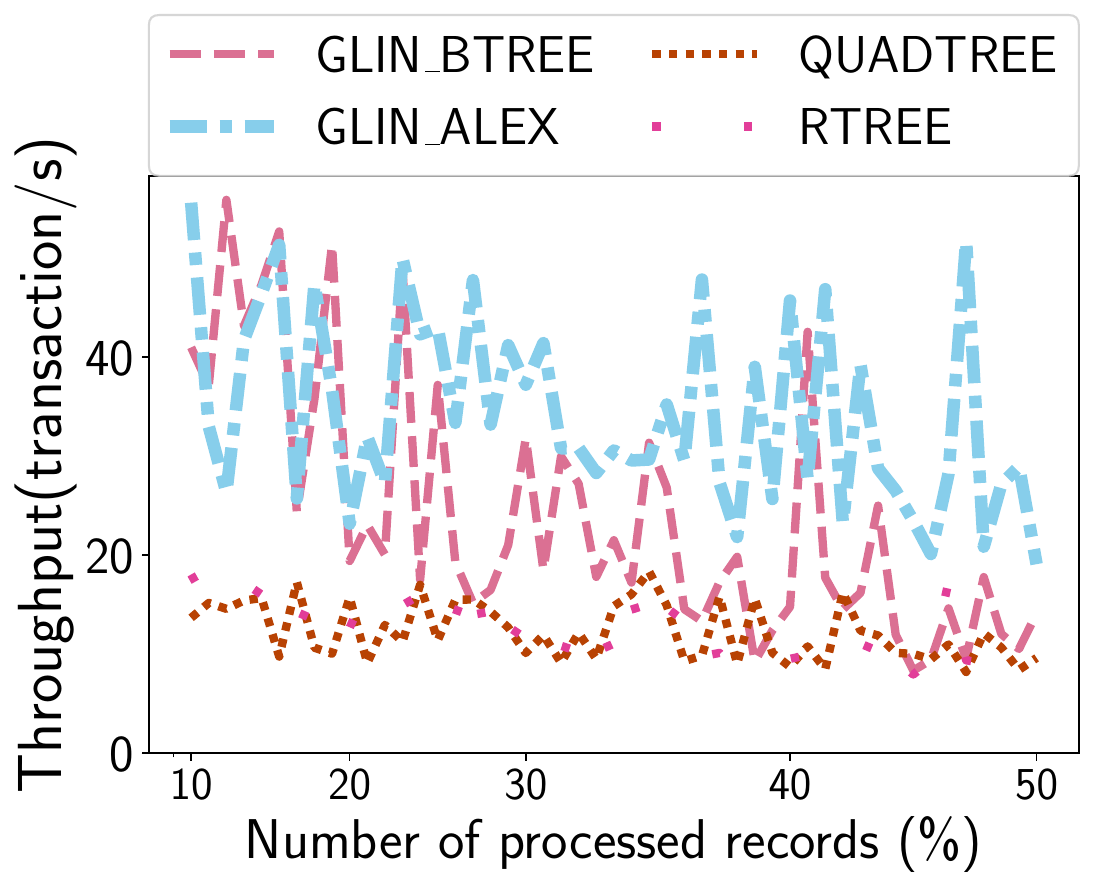}
		\subcaption{Write-intensive on Roads}
		\label{subfig:write-intensive-roads}	
	\end{minipage}
	
	\caption{Index performance on hybrid workloads}
	\label{fig:heavyread}
	\vspace{0pt}
\end{figure}

We define a transaction as (1) query: a spatial range query with $Intersects$ relationship at 1\% selectivity, or (2) insertion: insert 1\% new records into the indexes. We have two hybrid workloads: (1) read-intensive: 90\% of the transactions are queries and the other 10\% are insertion. (2) Write-intensive: 50\% of the transactions are queries and the rest are insertion. For each dataset, we first bulk-load 50\% of the entire dataset, and then start the workloads. We stop when the remaining 50\% data are inserted.

\textbf{Read-intensive workload.}
As depicted in Figure~\ref{subfig:read-intensive-roads}, the throughput of {\indextitle}-ALEX and {\indextitle}-BTREE initially surpasses that of Quad-Tree and R-Tree, but eventually aligns with the throughput of R-Tree and Quad-Tree towards the end. This behavior is expected, as {\indextitle}-BTREE requires more rebalancing as more data is inserted into the tree. Meanwhile, {\indextitle}-ALEX is able to consistently maintain higher throughput due to its learned index structure.

\textbf{Write-intensive workload.}
As shown in Figure~\ref{subfig:write-intensive-roads},  {\indextitle} outperforms Quad-Tree and R-Tree almost all the time. This matches our expectation because the insertion speed of {\indextitle} is much higher than that of Quad-Tree and R-Tree (see Figure~\ref{fig:realinsertion}). When we have a write-intensive workload, the overall performance of {\indextitle} is proven to be better.


\begin{thebibliography}{31}


\ifx \showCODEN    \undefined \def \showCODEN     #1{\unskip}     \fi
\ifx \showDOI      \undefined \def \showDOI       #1{#1}\fi
\ifx \showISBNx    \undefined \def \showISBNx     #1{\unskip}     \fi
\ifx \showISBNxiii \undefined \def \showISBNxiii  #1{\unskip}     \fi
\ifx \showISSN     \undefined \def \showISSN      #1{\unskip}     \fi
\ifx \showLCCN     \undefined \def \showLCCN      #1{\unskip}     \fi
\ifx \shownote     \undefined \def \shownote      #1{#1}          \fi
\ifx \showarticletitle \undefined \def \showarticletitle #1{#1}   \fi
\ifx \showURL      \undefined \def \showURL       {\relax}        \fi
\providecommand\bibfield[2]{#2}
\providecommand\bibinfo[2]{#2}
\providecommand\natexlab[1]{#1}
\providecommand\showeprint[2][]{arXiv:#2}

\bibitem[Athanassoulis and Ailamaki(2014)]%
        {AA14}
\bibfield{author}{\bibinfo{person}{Manos Athanassoulis} {and} \bibinfo{person}{Anastasia Ailamaki}.} \bibinfo{year}{2014}\natexlab{}.
\newblock \showarticletitle{BF-Tree: Approximate Tree Indexing}.
\newblock \bibinfo{journal}{\emph{PVLDB}} \bibinfo{volume}{7}, \bibinfo{number}{14} (\bibinfo{year}{2014}), \bibinfo{pages}{1881--1892}.
\newblock


\bibitem[Bentley(1975)]%
        {B75}
\bibfield{author}{\bibinfo{person}{Jon~Louis Bentley}.} \bibinfo{year}{1975}\natexlab{}.
\newblock \showarticletitle{Multidimensional Binary Search Trees Used for Associative Searching}.
\newblock \bibinfo{journal}{\emph{CACM}} \bibinfo{volume}{18}, \bibinfo{number}{9} (\bibinfo{year}{1975}), \bibinfo{pages}{509--517}.
\newblock


\bibitem[Bouros and Mamoulis(2019)]%
        {BM19}
\bibfield{author}{\bibinfo{person}{Panagiotis Bouros} {and} \bibinfo{person}{Nikos Mamoulis}.} \bibinfo{year}{2019}\natexlab{}.
\newblock \showarticletitle{Spatial joins: what's next?}
\newblock \bibinfo{journal}{\emph{{ACM} {SIGSPATIAL} Special}} \bibinfo{volume}{11}, \bibinfo{number}{1} (\bibinfo{year}{2019}), \bibinfo{pages}{13--21}.
\newblock


\bibitem[Davitkova et~al\mbox{.}(2020)]%
        {DMM+20}
\bibfield{author}{\bibinfo{person}{Angjela Davitkova}, \bibinfo{person}{Evica Milchevski}, {and} \bibinfo{person}{Sebastian Michel}.} \bibinfo{year}{2020}\natexlab{}.
\newblock \showarticletitle{The ML-Index: {A} Multidimensional, Learned Index for Point, Range, and Nearest-Neighbor Queries}. In \bibinfo{booktitle}{\emph{EDBT}}. \bibinfo{pages}{407--410}.
\newblock


\bibitem[degree-precision({[n.\,d.]})]%
        {degree-precision}
degree-precision \bibinfo{year}{[n.\,d.]}\natexlab{}.
\newblock \bibinfo{title}{Accuracy versus decimal places}.
\newblock \bibinfo{howpublished}{{\url{http://wiki.gis.com/wiki/index.php/Decimal_degrees}}}.
\newblock


\bibitem[Ding et~al\mbox{.}(2020)]%
        {DMY+20}
\bibfield{author}{\bibinfo{person}{Jialin Ding}, \bibinfo{person}{Umar~Farooq Minhas}, \bibinfo{person}{Jia Yu}, \bibinfo{person}{Chi Wang}, \bibinfo{person}{Jaeyoung Do}, \bibinfo{person}{Yinan Li}, \bibinfo{person}{Hantian Zhang}, \bibinfo{person}{Badrish Chandramouli}, \bibinfo{person}{Johannes Gehrke}, \bibinfo{person}{Donald Kossmann}, \bibinfo{person}{David~B. Lomet}, {and} \bibinfo{person}{Tim Kraska}.} \bibinfo{year}{2020}\natexlab{}.
\newblock \showarticletitle{{ALEX:} An Updatable Adaptive Learned Index}. In \bibinfo{booktitle}{\emph{SIGMOD}}. \bibinfo{pages}{969--984}.
\newblock


\bibitem[Eldawy and Mokbel(2015)]%
        {EM+15}
\bibfield{author}{\bibinfo{person}{Ahmed Eldawy} {and} \bibinfo{person}{Mohamed~F. Mokbel}.} \bibinfo{year}{2015}\natexlab{}.
\newblock \showarticletitle{{SpatialHadoop: {A} MapReduce Framework for Spatial Data}}. In \bibinfo{booktitle}{\emph{ICDE}}. \bibinfo{pages}{1352--1363}.
\newblock


\bibitem[geometries({[n.\,d.]})]%
        {geometries}
geometries \bibinfo{year}{[n.\,d.]}\natexlab{}.
\newblock \bibinfo{title}{{ISO/IEC 13249-3:2016 Information technology — Database languages — SQL multimedia and application packages — Part 3: Spatial}}.
\newblock
\newblock
\newblock
\shownote{\url{https://www.iso.org/standard/60343.html}}.


\bibitem[geoparquet({[n.\,d.]})]%
        {geoparquet}
geoparquet \bibinfo{year}{[n.\,d.]}\natexlab{}.
\newblock \bibinfo{title}{GeoParquet}.
\newblock \bibinfo{howpublished}{{\url{https://github.com/opengeospatial/geoparquet}}}.
\newblock


\bibitem[Guttman(1984)]%
        {G84}
\bibfield{author}{\bibinfo{person}{Antonin Guttman}.} \bibinfo{year}{1984}\natexlab{}.
\newblock \showarticletitle{R-Trees: {A} Dynamic Index Structure for Spatial Searching}. In \bibinfo{booktitle}{\emph{SIGMOD}}. \bibinfo{pages}{47--57}.
\newblock


\bibitem[Hentschel et~al\mbox{.}(2018)]%
        {HKI+18}
\bibfield{author}{\bibinfo{person}{Brian Hentschel}, \bibinfo{person}{Michael~S. Kester}, {and} \bibinfo{person}{Stratos Idreos}.} \bibinfo{year}{2018}\natexlab{}.
\newblock \showarticletitle{Column Sketches: {A} Scan Accelerator for Rapid and Robust Predicate Evaluation}. In \bibinfo{booktitle}{\emph{SIGMOD}}. \bibinfo{pages}{857--872}.
\newblock


\bibitem[Jagadish et~al\mbox{.}(2005)]%
        {JOT+05}
\bibfield{author}{\bibinfo{person}{H.~V. Jagadish}, \bibinfo{person}{Beng~Chin Ooi}, {and} \bibinfo{person}{Kian{-}Lee Tan}.} \bibinfo{year}{2005}\natexlab{}.
\newblock \showarticletitle{iDistance: An adaptive B\({}^{\mbox{+}}\)-tree based indexing method for nearest neighbor search}.
\newblock \bibinfo{journal}{\emph{TODS}} \bibinfo{volume}{30}, \bibinfo{number}{2} (\bibinfo{year}{2005}), \bibinfo{pages}{364--397}.
\newblock


\bibitem[Kipf et~al\mbox{.}(2020)]%
        {KMR+20}
\bibfield{author}{\bibinfo{person}{Andreas Kipf}, \bibinfo{person}{Ryan Marcus}, \bibinfo{person}{Alexander van Renen}, \bibinfo{person}{Mihail Stoian}, \bibinfo{person}{Alfons Kemper}, \bibinfo{person}{Tim Kraska}, {and} \bibinfo{person}{Thomas Neumann}.} \bibinfo{year}{2020}\natexlab{}.
\newblock \showarticletitle{RadixSpline: a single-pass learned index}. In \bibinfo{booktitle}{\emph{SIGMOD}}. \bibinfo{pages}{5:1--5:5}.
\newblock


\bibitem[Kraska et~al\mbox{.}(2018)]%
        {KBC+18}
\bibfield{author}{\bibinfo{person}{Tim Kraska}, \bibinfo{person}{Alex Beutel}, \bibinfo{person}{Ed~H. Chi}, \bibinfo{person}{Jeffrey Dean}, {and} \bibinfo{person}{Neoklis Polyzotis}.} \bibinfo{year}{2018}\natexlab{}.
\newblock \showarticletitle{The Case for Learned Index Structures}. In \bibinfo{booktitle}{\emph{SIGMOD}}. \bibinfo{pages}{489--504}.
\newblock


\bibitem[Lee et~al\mbox{.}(2007)]%
        {LZL+07}
\bibfield{author}{\bibinfo{person}{Ken C.~K. Lee}, \bibinfo{person}{Baihua Zheng}, \bibinfo{person}{Huajing Li}, {and} \bibinfo{person}{Wang{-}Chien Lee}.} \bibinfo{year}{2007}\natexlab{}.
\newblock \showarticletitle{Approaching the Skyline in {Z} Order}. In \bibinfo{booktitle}{\emph{VLDB}}. \bibinfo{pages}{279--290}.
\newblock


\bibitem[Li et~al\mbox{.}(2020)]%
        {LLZ+20}
\bibfield{author}{\bibinfo{person}{Pengfei Li}, \bibinfo{person}{Hua Lu}, \bibinfo{person}{Qian Zheng}, \bibinfo{person}{Long Yang}, {and} \bibinfo{person}{Gang Pan}.} \bibinfo{year}{2020}\natexlab{}.
\newblock \showarticletitle{{LISA:} {A} Learned Index Structure for Spatial Data}. In \bibinfo{booktitle}{\emph{SIGMOD}}. \bibinfo{pages}{2119--2133}.
\newblock


\bibitem[libmorton({[n.\,d.]})]%
        {libmorton}
libmorton \bibinfo{year}{[n.\,d.]}\natexlab{}.
\newblock \bibinfo{title}{Libmorton Library}.
\newblock \bibinfo{howpublished}{{\url{https://github.com/Forceflow/libmorton}}}.
\newblock


\bibitem[Nathan et~al\mbox{.}(2020)]%
        {NDA+20}
\bibfield{author}{\bibinfo{person}{Vikram Nathan}, \bibinfo{person}{Jialin Ding}, \bibinfo{person}{Mohammad Alizadeh}, {and} \bibinfo{person}{Tim Kraska}.} \bibinfo{year}{2020}\natexlab{}.
\newblock \showarticletitle{Learning Multi-Dimensional Indexes}. In \bibinfo{booktitle}{\emph{SIGMOD}}. \bibinfo{pages}{985--1000}.
\newblock


\bibitem[Olma et~al\mbox{.}(2017)]%
        {OTH+17}
\bibfield{author}{\bibinfo{person}{Matthaios Olma}, \bibinfo{person}{Farhan Tauheed}, \bibinfo{person}{Thomas Heinis}, {and} \bibinfo{person}{Anastasia Ailamaki}.} \bibinfo{year}{2017}\natexlab{}.
\newblock \showarticletitle{{BLOCK:} Efficient Execution of Spatial Range Queries in Main-Memory}. In \bibinfo{booktitle}{\emph{SSDBM}}. \bibinfo{pages}{15:1--15:12}.
\newblock


\bibitem[OSM({[n.\,d.]})]%
        {OSM}
OSM \bibinfo{year}{[n.\,d.]}\natexlab{}.
\newblock \bibinfo{title}{{OpenStreetMap}}.
\newblock
\newblock
\newblock
\shownote{\url{http://www.openstreetmap.org/}}.


\bibitem[parquet({[n.\,d.]})]%
        {parquet}
parquet \bibinfo{year}{[n.\,d.]}\natexlab{}.
\newblock \bibinfo{title}{Apache Parquet}.
\newblock \bibinfo{howpublished}{{\url{https://parquet.apache.org/}}}.
\newblock


\bibitem[Qi et~al\mbox{.}(2020)]%
        {QLJ+20}
\bibfield{author}{\bibinfo{person}{Jianzhong Qi}, \bibinfo{person}{Guanli Liu}, \bibinfo{person}{Christian~S. Jensen}, {and} \bibinfo{person}{Lars Kulik}.} \bibinfo{year}{2020}\natexlab{}.
\newblock \showarticletitle{Effectively Learning Spatial Indices}.
\newblock \bibinfo{journal}{\emph{PVLDB}} \bibinfo{volume}{13}, \bibinfo{number}{11} (\bibinfo{year}{2020}), \bibinfo{pages}{2341--2354}.
\newblock


\bibitem[Samet(1984)]%
        {S84}
\bibfield{author}{\bibinfo{person}{Hanan Samet}.} \bibinfo{year}{1984}\natexlab{}.
\newblock \showarticletitle{The Quadtree and Related Hierarchical Data Structures}.
\newblock \bibinfo{journal}{\emph{CSUR}} \bibinfo{volume}{16}, \bibinfo{number}{2} (\bibinfo{year}{1984}), \bibinfo{pages}{187--260}.
\newblock


\bibitem[Sidirourgos and Kersten(2013)]%
        {SK+13}
\bibfield{author}{\bibinfo{person}{Lefteris Sidirourgos} {and} \bibinfo{person}{Martin~L. Kersten}.} \bibinfo{year}{2013}\natexlab{}.
\newblock \showarticletitle{Column imprints: a secondary index structure}. In \bibinfo{booktitle}{\emph{SIGMOD}}. \bibinfo{pages}{893--904}.
\newblock


\bibitem[tiger({[n.\,d.]})]%
        {Tiger}
tiger \bibinfo{year}{[n.\,d.]}\natexlab{}.
\newblock \bibinfo{title}{{TIGER/Line files}}.
\newblock
\newblock
\newblock
\shownote{\url{http://www.census.gov/geo/www/tiger/}}.


\bibitem[Tsitsigkos et~al\mbox{.}(2021)]%
        {TLB+21}
\bibfield{author}{\bibinfo{person}{Dimitrios Tsitsigkos}, \bibinfo{person}{Konstantinos Lampropoulos}, \bibinfo{person}{Panagiotis Bouros}, \bibinfo{person}{Nikos Mamoulis}, {and} \bibinfo{person}{Manolis Terrovitis}.} \bibinfo{year}{2021}\natexlab{}.
\newblock \showarticletitle{A Two-layer Partitioning for Non-point Spatial Data}. In \bibinfo{booktitle}{\emph{ICDE}}. \bibinfo{pages}{1787--1798}.
\newblock


\bibitem[Wang et~al\mbox{.}(2019)]%
        {WFX+19}
\bibfield{author}{\bibinfo{person}{Haixin Wang}, \bibinfo{person}{Xiaoyi Fu}, \bibinfo{person}{Jianliang Xu}, {and} \bibinfo{person}{Hua Lu}.} \bibinfo{year}{2019}\natexlab{}.
\newblock \showarticletitle{Learned Index for Spatial Queries}. In \bibinfo{booktitle}{\emph{MDM}}. \bibinfo{pages}{569--574}.
\newblock


\bibitem[Wu et~al\mbox{.}(2019)]%
        {WYT+19}
\bibfield{author}{\bibinfo{person}{Yingjun Wu}, \bibinfo{person}{Jia Yu}, \bibinfo{person}{Yuanyuan Tian}, \bibinfo{person}{Richard Sidle}, {and} \bibinfo{person}{Ronald Barber}.} \bibinfo{year}{2019}\natexlab{}.
\newblock \showarticletitle{Designing Succinct Secondary Indexing Mechanism by Exploiting Column Correlations}. In \bibinfo{booktitle}{\emph{SIGMOD}}. \bibinfo{pages}{1223--1240}.
\newblock


\bibitem[Yu and Sarwat(2016)]%
        {YS16}
\bibfield{author}{\bibinfo{person}{Jia Yu} {and} \bibinfo{person}{Mohamed Sarwat}.} \bibinfo{year}{2016}\natexlab{}.
\newblock \showarticletitle{Two Birds, One Stone: {A} Fast, yet Lightweight, Indexing Scheme for Modern Database Systems}.
\newblock \bibinfo{journal}{\emph{PVLDB}} \bibinfo{volume}{10}, \bibinfo{number}{4} (\bibinfo{year}{2016}), \bibinfo{pages}{385--396}.
\newblock


\bibitem[Yu and Sarwat(2017)]%
        {YS17}
\bibfield{author}{\bibinfo{person}{Jia Yu} {and} \bibinfo{person}{Mohamed Sarwat}.} \bibinfo{year}{2017}\natexlab{}.
\newblock \showarticletitle{Indexing the Pickup and Drop-Off Locations of {NYC} Taxi Trips in PostgreSQL - Lessons from the Road}. In \bibinfo{booktitle}{\emph{SSTD}} \emph{(\bibinfo{series}{Lecture Notes in Computer Science}, Vol.~\bibinfo{volume}{10411})}. \bibinfo{pages}{145--162}.
\newblock


\bibitem[Yu et~al\mbox{.}(2019)]%
        {YZS19}
\bibfield{author}{\bibinfo{person}{Jia Yu}, \bibinfo{person}{Zongsi Zhang}, {and} \bibinfo{person}{Mohamed Sarwat}.} \bibinfo{year}{2019}\natexlab{}.
\newblock \showarticletitle{Spatial data management in apache spark: the GeoSpark perspective and beyond}.
\newblock \bibinfo{journal}{\emph{GeoInformatica}} \bibinfo{volume}{23}, \bibinfo{number}{1} (\bibinfo{year}{2019}), \bibinfo{pages}{37--78}.
\newblock


\end{thebibliography}
\end{document}